\documentclass[useAMS,usenatbib]{mn2e}
\usepackage{lscape}
\usepackage{amsmath}
\usepackage{latexsym}
\usepackage{graphicx}
\usepackage{amssymb}

\newcommand{\lyap}{Ly$\alpha$}
\newcommand{\lya}{Ly$\alpha$ }

\newcommand{\zz}{$z\sim2$ }
\newcommand{\zzzp}{$z\sim3$}
\newcommand{\zzz}{$z\sim3$ }
\newcommand{\kmsp}{km s$^{-1}$}
\newcommand{\kms}{km s$^{-1}$ }

\newcommand{\h}{$h^{-1}$ }
\newcommand{\kpcp}{$h^{-1}$kpc}
\newcommand{\kpc}{$h^{-1}$kpc }
\newcommand{\mpcp}{$h^{-1}$Mpc}
\newcommand{\mpc}{$h^{-1}$Mpc }
\newcommand{\scpp}{SSCpairs}
\newcommand{\scp}{SSCpairs }

\title[Close galaxy pairs at $z\sim3$]{Lyman break galaxy close and
  interacting pairs at $z\sim3$}

\author[Cooke et al.]{Jeff Cooke$^{1}$\thanks{E-mail: cooke@uci.edu
(JC)
}\thanks{Gary McCue Postdoctoral Fellow}, Joel C. Berrier$^{1}$,
Elizabeth J. Barton$^{1}$, James S. Bullock$^{1}$, \newauthor and
Arthur M.  Wolfe$^{2}$ \\ $^{1}$Department of Physics \& Astronomy,
Center for Cosmology, University of California, Irvine, Irvine, CA,
92697, USA\\ $^{2}$Department of Physics, Center for Astophysics and
Space Sciences, University of California, San Diego, La Jolla, CA,
92093, USA}

\begin{document}

\date{Accepted 2009 December 07. Received 2009 December 05; in
original form 2009 October 31}
\pagerange{\pageref{firstpage}--\pageref{lastpage}} \pubyear{0000}

\maketitle

\label{firstpage}
 
\begin{abstract}
Investigations of interacting and merging galaxies at high redshift
are vital to our understanding of their formation and evolution.  To
date, the identification of interactions at \zzz and above has relied
on rest-frame ultraviolet morphological parameters.  Here, we present
five serendipitous spectroscopic \zzz Lyman break galaxy (LBG) pairs
with projected proper separations $<15$ \kpc in our survey of nine
separate deep Keck fields.  The data consist of $140$ of our highest
signal-to-noise ratio LBG spectra and $\sim500$ of our most confident
colour-selected LBGs.  We show that the pairs are composed of two
distinct close and/or interacting LBGs from a detailed analysis of the
rest-frame ultraviolet 1-D and 2-D spectra and the deep broadband
images.  In addition, we show that the number and separation
distribution of the pairs is expected from (1) the two-point angular
correlation function when applied to the LBG pair separation
distribution in our survey and $\sim2500$ colour-selected LBGs from
the literature and (2) an analysis of a carefully matched
high-resolution hybrid numerical and analytical cosmological
simulation.  Because the spectroscopic slitlets have random
orientations with respect to the close pairs on the sky, the
serendipitous pairs provide an unbiased sampling of the underlying
close pair fraction.  Finally, we discover two \lya emitters (LAEs) in
our slitlets and find that they reside within $50$ projected \kpc of
the spectroscopic LBGs.  In this work, we uncover a strong
relationship between \lya emission and pair separation.  All confirmed
and all candidate LBG pairs with separations of $\le15$ projected \kpc
exhibit \lya in emission and we find an indication of an overabundance
of \lya emission in pairs with $\le50$ projected \kpc separations.
This relationship suggests a picture in which a measurable fraction of
the \lya emission of LBGs, and potentially LAEs, is generated via
interaction mechanisms such as triggered star formation and the
dispersal of obscuring gas and dust.  As a result, serendipitous
spectroscopic close pairs provide a unique means to help identify and
study high-redshift galaxy interactions using ground-based optical
data.
\end{abstract}

\begin{keywords}
methods: observational -- galaxies: evolution -- galaxies: formation
-- galaxies: high-redshift -- galaxies: interactions
\end{keywords}

\section{INTRODUCTION}\label{intro}

A fundamental prediction of LCDM models of structure formation is the
hierarchical build up of galaxies via mergers with a rate that
declines over time \citep[e.g.,][and references therein]{cole00}.
Observations of this process at low and high redshift are necessary to
test these predictions, assess the effects of interactions, and to
help clarify our current understanding of galaxy assembly.  Close
interactions and mergers act to trigger star formation and contribute
to the accretion of matter and the morphological transformation of
galaxies over time \citep[e.g.,][and many
others]{larson78,keel85,mihos96,barton00,barton03,steinmetz02,scan03}.

In practice, however, identifying and studying mergers and
interactions is difficult, largely because definitions vary, tidal
features and interaction indicators can be faint, and morphologies are
difficult to quantify.  At higher redshifts ($z\gtrsim1$), this is
more challenging as galaxies become fainter and less resolved.
Although nonparametric methods have been developed to characterise the
morphology of high-redshift systems \citep{conselice03,lotz06,law07},
such analyses are best done using the superior efficiency and
resolution of optical data but, as a result, contribute an additional
complication of linking the observed clumpy and chaotic rest-frame
ultraviolet (UV) to the better understood rest-frame optical
morphology.

The Lyman break galaxies (LBGs) detected by the Lyman break
colour-selection technique \citep{s96} are a well-studied population
of galaxies at high-redshift.  Deep {\it Hubble Space Telescope} (HST)
images of $z>2$ LBGs in surveys such as GOODS, COSMOS, and the UDF
show a wide range of rest-frame UV morphologies -- from single compact
or diffuse systems to multiple compact star forming components with,
or without, diffuse emission \citep[e.g,][]{law07,rafelski09}.  The
latter morphology has been interpreted as either star forming regions
within a single coalescing galaxy and/or the result of interactions or
mergers.

Because of the faint nature of \zzz LBGs ($L*$ corresponds to
m$_R\sim24.5$), low-resolution spectroscopy can only achieve a
signal-to-noise ratio (S/N) of a few with 8m-class telescopes for
galaxies with m$_R\lesssim25.5$ using reasonable integration times.
This limitation prohibits detailed investigations of individual LBGs.
Except for the very brightest cases, detailed studies require
composite spectra and/or rare gravitationally lensed systems.
Composite spectral analysis \citep[e.g.,][]{aes03} has provided great
insight into the average properties of LBGs but dilutes the features
and relationships particular to individual systems.  Gravitationally
lensed LBG systems provide magnified flux that boosts the S/N of
individual systems, but the total number of these systems is small,
their reconstruction analyses are subject to model constraints, and
their chance discoveries do not allow a study at desired redshifts.

Recently, infrared integral field observations using adaptive optics
have been very informative on the kinematics of a sample of \zz LBGs
\citep[e.g.,][]{fs06,law07,shapiro08,wright09} from prominent
rest-frame optical features.  Although this approach has been
effective on relatively luminous LBGs, it would be prohibitively time
consuming to compile a statistical sample of typical (including faint)
systems using existing instruments.  Moreover, this method becomes
increasingly difficult for higher redshift populations as key features
are shifted out of the range of wavelength sensitivity.  As a result
of the above limitations, a clear and consistent interpretation of the
morphological and spectroscopic effect of mergers and interactions of
typical LBGs at $z\gtrsim3$ via conventional approaches lay beyond the
sensitivity thresholds and capabilities of current ground-based
AO-assisted or space-based observations.

A standard method to identify a large number of merger and interacting
galaxy candidates at low- to intermediate redshift and to avoid many
of the difficulties in morphological classification and interpretation
is to measure the galaxy close pair fraction
\citep[e.g.,][]{patton00,patton02,lin04,berrier06,lin07,kart07}.
Galaxy pairs with observed separations of $\le50$ \kpc and velocity
offsets of $\pm500$ \kms are likely to interact and/or merge and can
be more easily and consistently observed and quantified.  Moreover,
galaxies found in close pairs at high redshift have a higher
likelihood of an imminent merger than at low redshift based in part on
the fact that there are fewer galaxy clusters
\citep[e.g.,][]{berrier06}.  As a result, close pairs at high redshift
provide a means to identify galaxies that are interacting or are
destined to interact on a short timescale and a strong constraint on
the merger rate.  Finally, the detailed behaviour of close pairs can
be tracked in matched high resolution cosmological simulations to
glean information beyond that of the observations.

Although the expected number of LBGs in close pairs can be inferred
photometrically from the observed angular correlation function,
spectroscopy is necessary to help differentiate actual close pairs
from false pairs that have small projected separations on the sky.  In
order to gather a large number of LBG spectra, nearly all surveys to
date have utilised optical low-resolution multi-object spectroscopy
(MOS).  This method uses slitmasks to place slitlets at the locations
of selected targets but cannot acquire spectroscopy for objects with
very small projected separations on the sky or that conflict in the
dispersion direction.  While MOS is efficient in acquiring a
statistical sample of typical LBGs, this shortcoming makes it
difficult or impossible for a single slitmask to acquire information
for many of the galaxies closely spaced on the sky.  A programme
employing a number of overlapping slitmask observations of the same
area of the sky could, in principle, resolve this issue.  However, we
show below that not all close galaxy pairs at $z\sim3$ are easily
identifiable from their imaging alone and would introduce systematics
in such a time expensive endeavour.

A different technique, which we introduce here, allows existing MOS
datasets to be used to study close and interacting pairs at high
redshift.  The density and clustering behaviour of LBGs predict that a
small fraction of close pairs are expected to fall serendipitously
into the area probed by the MOS slitlets.  Moreover, because the
slitlets in conventional LBG surveys have random orientations with
respect to the serendipitous close pairs on the sky, they produce
samples unbiased by selection.  In this manner, the serendipitous
pairs reflect the underlying close and interacting LBG pair behaviour
and effectively circumvent the MOS mechanical constraints.

We analyse a conventionally acquired nine-field deep Keck imaging and
spectroscopic survey for $z\sim3$ LBGs and find five serendipitous
spectroscopic close pairs.  We use 509 colour-selected and 140
spectroscopically confirmed LBGs, the large \zzz photometric LBG
dataset of \citet{s03}, and a high-resolution cosmological simulation
to investigate the distribution of close pair separations and the
expectation of the serendipitous pairs.  We present several lines of
evidence from a detailed analysis of the imaging and spectroscopic
data that support the close pair/interacting nature of the
serendipitous detections.  We show that the properties of this unique
sample, in turn, provide insight into the rest-frame UV properties of
the LBG population that are otherwise inaccessible and a means to help
identify high-redshift interactions in ground-based data.

This paper is organised as follows: We describe relevant information
regarding the data and data acquisition in \S\ref{obs}. We present the
serendipitous spectroscopic interacting pairs, investigate their
rest-frame UV behaviour, and compare the results to precedence in the
literature in \S\ref{SCpairs}.  Close pair observations and
measurements are discussed in \S\ref{close_pairs} and the results from
the simulation analysis in \S\ref{sim}.  The two \lya emitters
discovered in our sample are presented in \S\ref{lae} and we address
the observed relationship between \lya in emission and close pair
separation in \S\ref{ecp}.  Finally in \S\ref{sum} we provide a
summary.  All magnitudes are in the AB \citep{f96} magnitude system
unless otherwise noted.  We assume an
$\Omega_M=0.3,\Omega_\Lambda=0.7,h=1.0$ cosmology throughout and list
close pair separations in physical (proper) units.


\section{OBSERVATIONS}\label{obs}

We use the \zzz LBG survey of \citet[][hereafter C05]{c05} for the
observational analysis in this paper.  The survey consists of deep
imaging and spectroscopy of nine widely-separated fields covering
$\sim460$ arcmin$^2$.  Seven fields were imaged using the Low
Resolution Imaging Spectrometer \citep[LRIS;][]{o95,mccarthy98} on the
10-m Keck I telescope with two fields imaged using the Carnegie
Observatories Spectrograph and Multi-Object Imaging Camera
\citep[COSMIC;][]{k98} mounted on the 5-m Hale telescope at the
Palomar Observatory.  We obtained follow-up spectroscopy of all nine
fields using LRIS.  The multiple fields suppress the effects of cosmic
variance and provide $\sim800$ colour-selected and $\sim200$
spectroscopically identified LBGs for statistical analysis.  Details
of the observations, data reduction, and LBG analysis can be found in
C05 and \citet{c06}.  Below, we briefly discuss a few relevant
specifics of the data.

The $z\sim3$ LBG candidates in our survey are selected by their
$u'$BVRI colours.  Although the deep images can select LBGs to fainter
magnitudes, we restrict our analysis in this paper to LBGs with
m$_R\le25.5$ in order to define a sample that has (1) photometry
accurate to $\le0.2$ mag uncertainty in all filters that probe the
rest-frame UV continua, (2) follow-up spectroscopy, (3) defined
photometric selection functions, and (4) a measured spatial
correlation function.

\begin{figure}
\begin{center}
\scalebox{0.56}[0.56]{\includegraphics{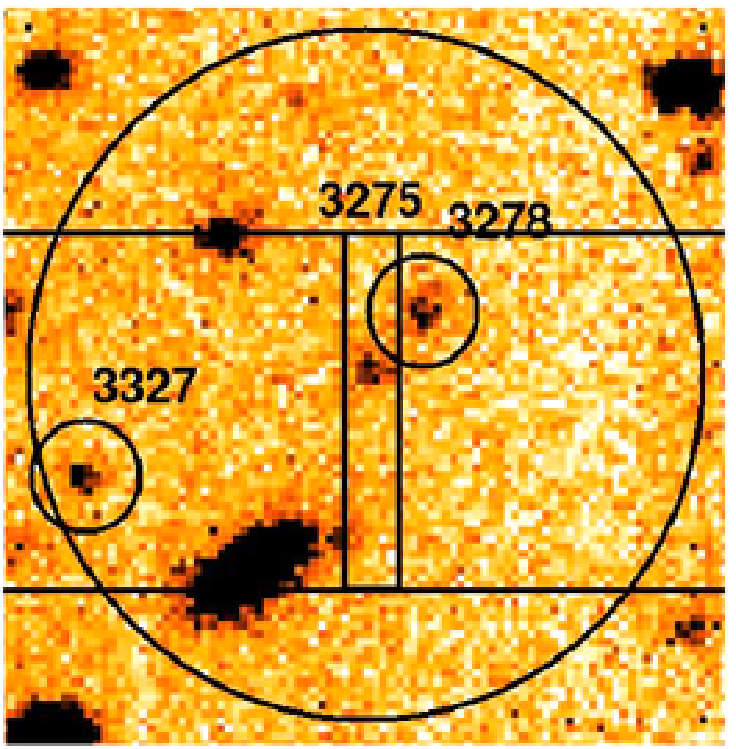}}
\scalebox{0.56}[0.56]{\includegraphics{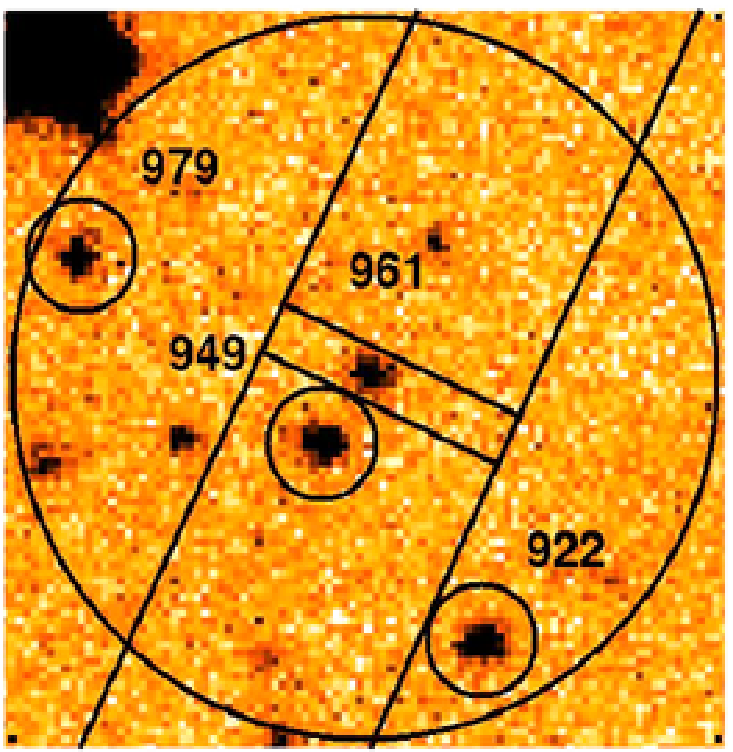}}
\caption
{\small Images of spectroscopic LBG close pair candidates ($\sim20''$
  on a side) centred on a spectroscopically confirmed $z\sim3$ galaxy.
  {\bf Left:} LBG 1057-3275, labelled 3275, at $z=3.014$ with
  colour-selected close pair galaxies 1057-3278 and 1057-3327 circled
  and labelled similarly.  {\bf Right:} LBG 1643-961 at $z=3.412$ with
  the colour-selected close pair candidates 1643-979, 1643-949, and
  1643-922.  This figure illustrates how the physical constraints of
  multi-object slitmasks can make it difficult to acquire spectroscopy
  of close galaxy pairs.  The central rectangle replicates the
  spectroscopic slitlet dimensions and orientation for that particular
  observation.  The larger outer circle represents $50$ \kpc physical
  radius from the central galaxy.  The region bounded by the lines
  that extend from the upper and lower edges of the slitlet indicate
  the dispersion direction that is inaccessible for that observation.
  For these particular observations, we were unable to obtain
  spectroscopy for candidates 1643-949, 1057-3278, and 1057-3327.
  Other constraints, such as the allowance of space between slitlets
  necessary for milling purposes, sufficient slitlet lengths for
  proper sky subtraction either side of the targeted object, and the
  presence of higher priority candidates in the dispersion direction
  (determined in the original survey) prevented the spectroscopic
  acquisition of close pair candidates such as 1643-922.}
\label{phot_pairs}
\end{center}
\end{figure}

The spectroscopic component of our survey follows a conventional
approach to gather a large number of \zzz LBG spectra using MOS
slitmasks.  The survey was not originally designed to study
interacting LBGs or to measure close spectroscopic pairs.  The
physical constraints of MOS slitmasks can prevent acquisition of
galaxies that are closely spaced on the sky or those that fall in the
dispersion direction of the targeted observations.  This situation is
illustrated in Figure~\ref{phot_pairs} for two spectroscopically
confirmed LBGs and associated colour-selected close pair candidates
from our survey.  However, because each MOS slitlet covers a
non-negligible area of the sky ($\sim18$ arcsec$^2$ on average),
occasionally a close pair will have a small enough separation or
coincidental near alignment to be detected in a slitlet that was
originally intended for a single photometrically identified
colour-selected LBG candidate.

The MOS slitlets in our survey were centred on the LBG candidates with
position angles near the parallactic angle of the observations to
minimise wavelength-dependent flux loss caused by atmospheric
dispersion.  As a result, this is an unbiased spectroscopic survey for
the purposes discussed here because the data were acquired randomly
with respect to the orientation of close pairs on the sky.  Any
detected close pairs in the slitlets are indeed serendipitous and
directly reflect the true underlying distribution of close pairs on
the scales probed by the dimensions of our slitlets.

The seeing in the images used for colour selection of the LBG sample
ranged from $0.''6-1.''5$ arcsec FWHM, with those having $0.''6-1.''0$
FWHM used for pair analysis.  In addition, all spectroscopy was
acquired under good conditions ($0.''8-1.''2$ arcsec seeing FWHM).
The slitlets were milled to have $1.''0-1.''5$ widths, with $6''$
minimum and $\sim14''$ average lengths.  With knowledge that \zzz LBGs
are near-point sources in ground-based observations, we designed the
widths of our MOS slitlets $\sim0.''4$ larger than the expected seeing
FWHM to gather the maximum flux and to compensate for
$\sim0.''2-0.''3$ astrometric errors across the focal plane.  As a
result, the spectral resolution for a given LBG was typically
determined by the seeing and is $\sim400$ \kms FWHM for most
observations.


\section{SERENDIPITOUS SPECTROSCOPIC CLOSE PAIRS}
\label{SCpairs}

In the full spectroscopic dataset, we discover LBGs that exhibit
unexpected double \lya emission in the 2-D spectra with double or
overlapping continua.  These data form distinct double \lya emission
peaks in the merged 1-D spectra.  In addition, we find that these
objects show double morphology (two measurably separated flux peaks)
in the broadband images.  Furthermore, we find that the \lya emission
and continua in the 2-D spectra have a one-to-one correspondence in
the spatial direction when compared to the separations of the flux
peaks in the images.  We term these objects ``serendipitous
spectroscopic close pairs'' (\scpp).

\begin{figure}
\begin{center}
\scalebox{0.56}[0.56]{\includegraphics{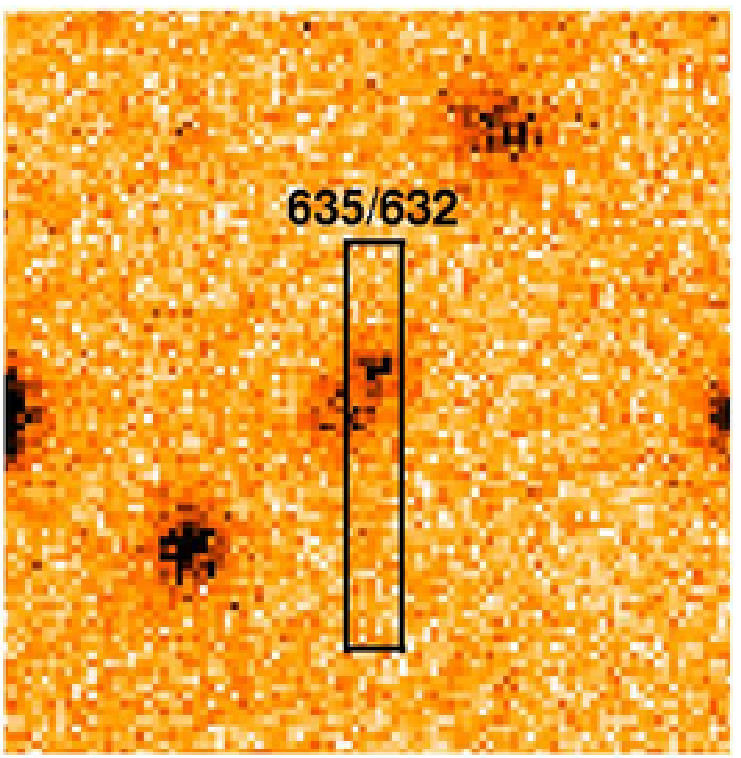}}
\scalebox{0.56}[0.56]{\includegraphics{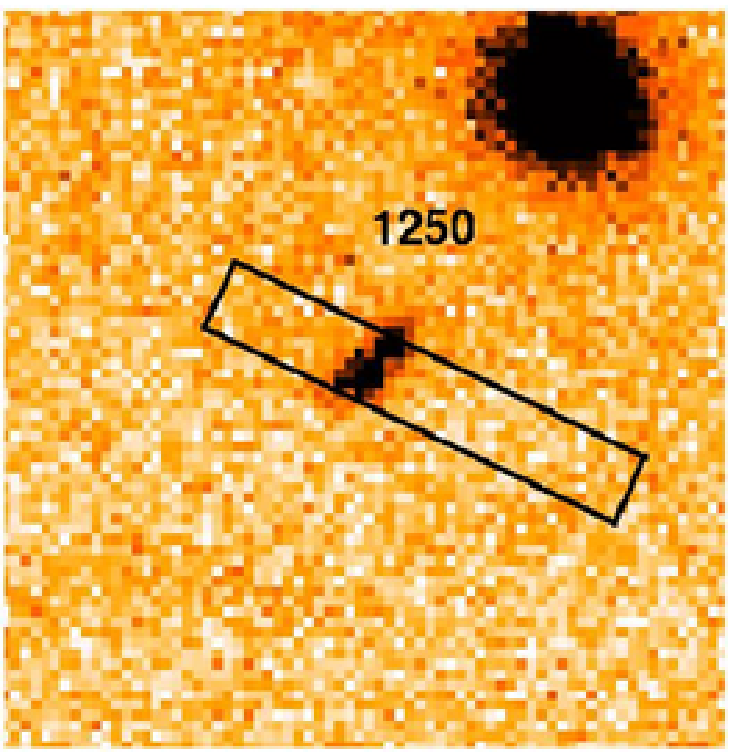}}
\scalebox{0.56}[0.84]{\includegraphics{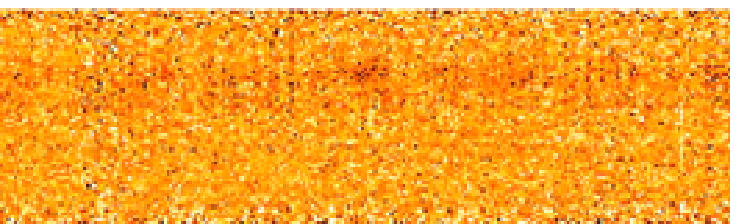}}
\scalebox{0.56}[0.84]{\includegraphics{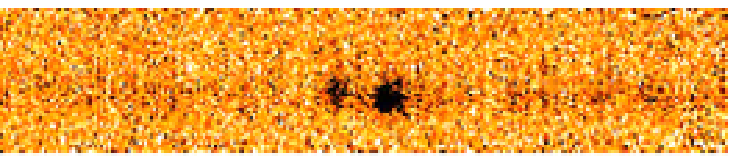}}
\scalebox{0.56}[0.56]{\includegraphics{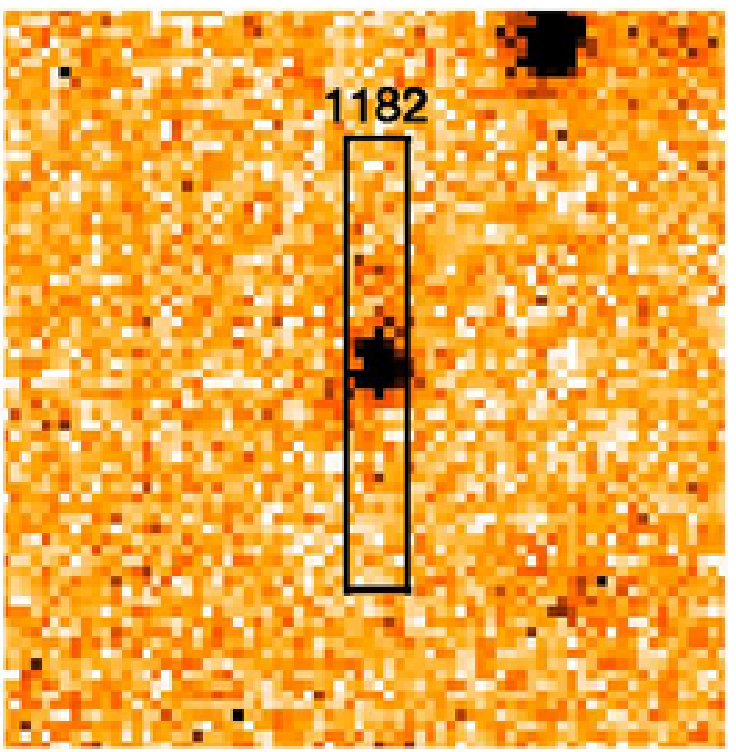}}
\scalebox{0.56}[0.56]{\includegraphics{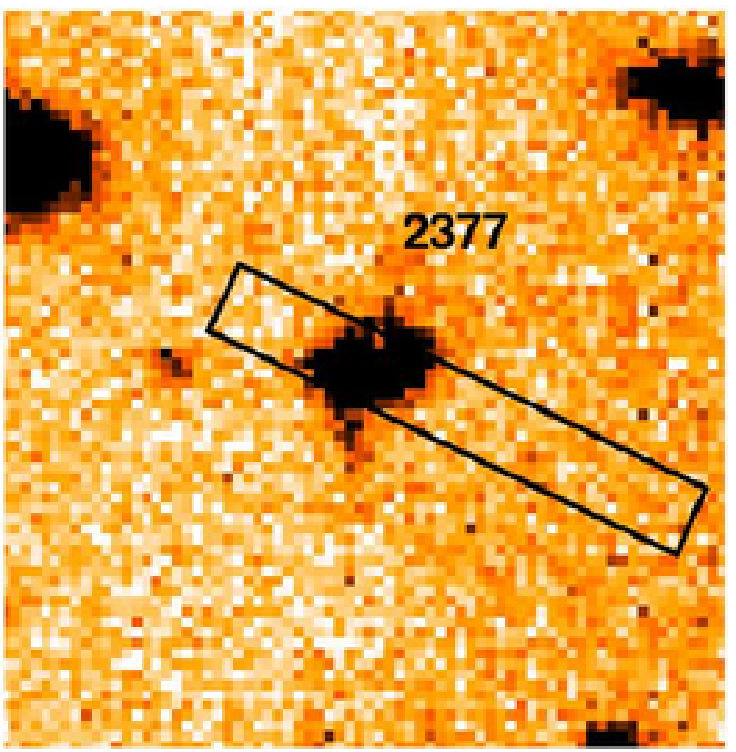}}
\scalebox{0.56}[0.84]{\includegraphics{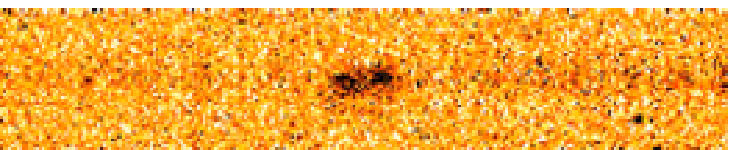}}
\scalebox{0.56}[0.84]{\includegraphics{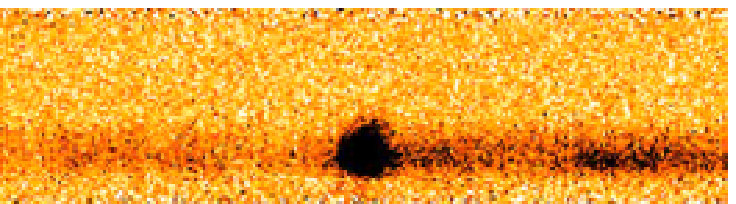}}
\scalebox{0.566}[0.56]{\includegraphics{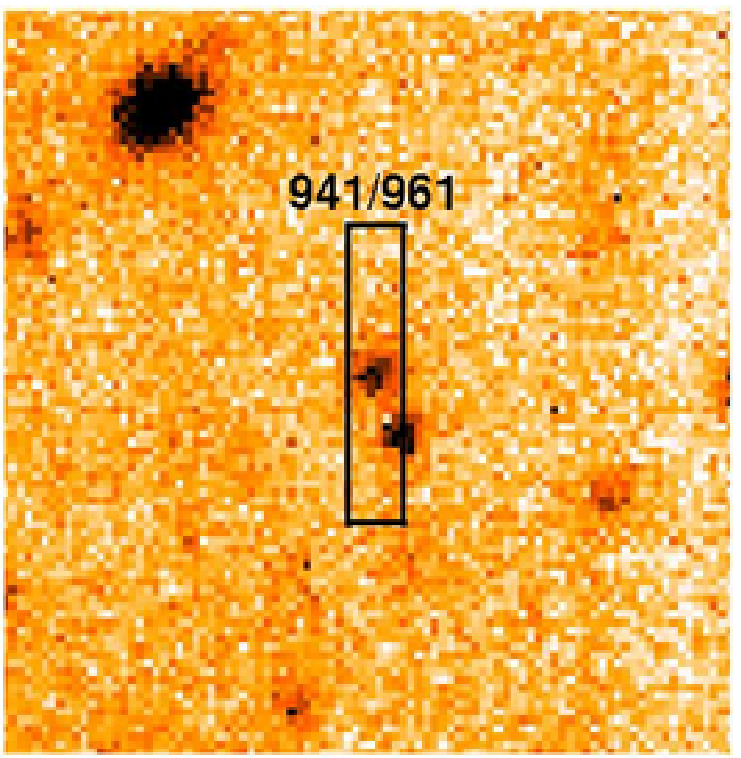}}
\scalebox{0.58}[0.84]{\includegraphics{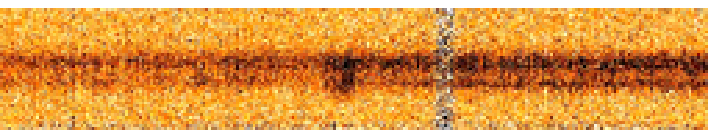}}
\caption
{\small Ground-based R-band images and 2-D spectra of the five
  spectroscopic \zzz LBG close pairs, the \scpp.  Each image is
  $\sim15''$ on a side and the labelled rectangle replicates the
  slitlet dimensions and orientation for that particular observation.
  Directly below each image is a section of the corresponding 2-D
  spectrum centred on the \lya feature.  {\bf Top:} \scp 0056-0635
  (0056-0635 and 0056-0632) and 0336-1250.  {\bf Centre:} \scp
  1013-1182 and 1643-2377.  {\bf Bottom:} \scp 0957-941 (0957-0941 and
  0957-0961).  The \scp are named after the original targeted LBG.
  Objects 0056-0632 and 0957-0961 are detected as separate
  colour-selected LBGs but are referred to as 0056-0635 and 0957-0941
  for consistency.  R-band images probe $z\sim3$ LBG UV continua near
  rest-frame $\sim1700$\AA.  Although faint, the 2-D spectrum of
  0056-0635 shows distinct continua and \lya emission for objects
  0056-0635 and 0056-0632.}
\label{Rband_2D}
\end{center}
\end{figure}

A systematic analysis the images, 2-D spectra, and 1-D spectra of 140
of our highest signal-to-noise ratio (S/N) spectroscopically confirmed
$z\sim3$ LBGs finds five such systems that show compelling evidence
for more than one identified major component.  We discuss the observed
properties of the five \scp in the following sections.  Because the
separations for some of the \scp cause them to be unresolved by the
source extraction routine {\it SExtractor} \citep{ba96} in the
ground-based images, we adopt a naming convention that uses the
R.A. of the targeted field and the identification number assigned by
{\it SExtractor} for the originally targeted LBG.  For the \scp that
were detected as two distinct LBGs by {\it SExtractor}, we label the
pair using the R.A. of the field and the identification number of the
targeted LBG only.  Figure~\ref{Rband_2D} displays the ground-based
LRIS R-band images and 2-D spectra of these systems.  Object 1643-2377
received follow-up higher S/N, higher resolution longslit spectroscopy
using LRIS \citep{c08}.  These data provide valuable insight into LBG
close pair properties and are used below to help confirm our analysis
methods of the low-resolution, low S/N SSCpair spectra.  For the
results presented here, including those from the simulation and close
pair analysis, we use the parameters of the low-resolution data of
1643-2377 for consistency.

\begin{figure}
\begin{center}
\scalebox{0.55}[0.55]{\includegraphics{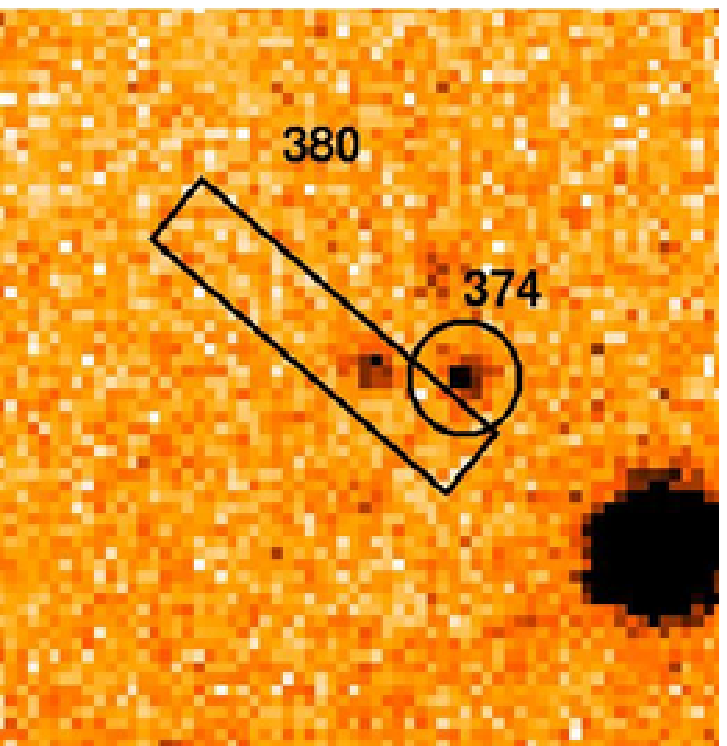}}
\scalebox{0.55}[0.55]{\includegraphics{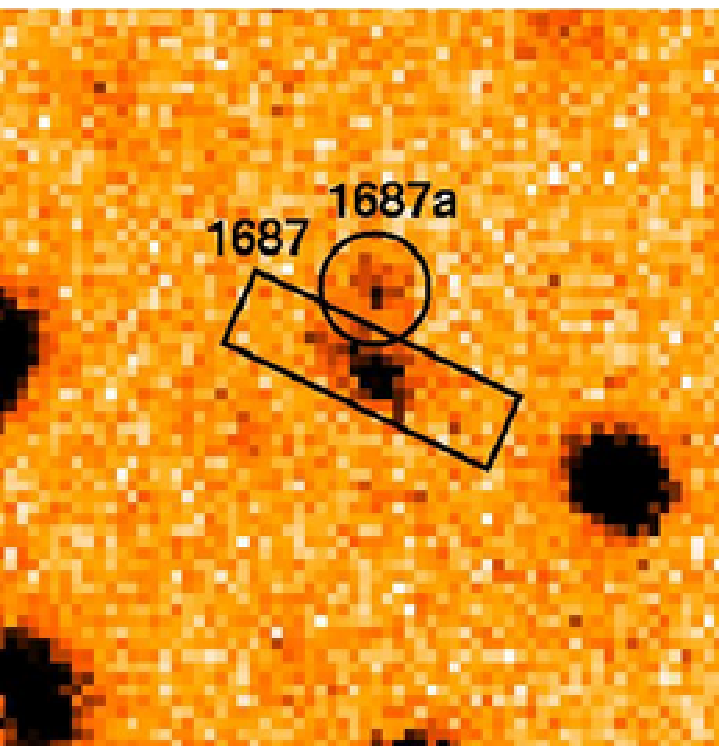}}
\scalebox{0.55}[0.55]{\includegraphics{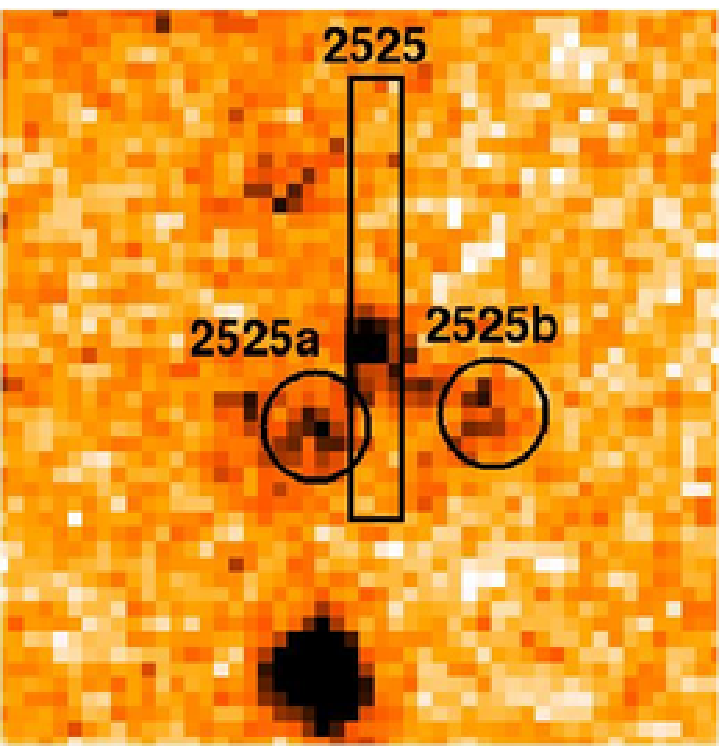}}
\scalebox{0.55}[0.55]{\includegraphics{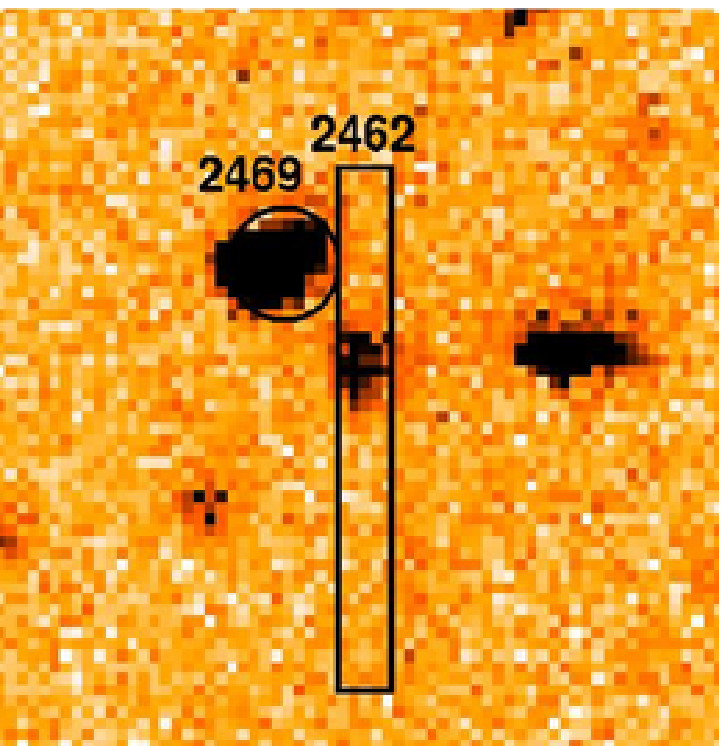}}
\caption
{\small Ground-based R-band images and 2-D spectra of four \zzz LBG
  $<15$ \kpc close pair candidates.  These four candidates, and the
  two candidates shown in Figure~\ref{phot_pairs}, were not included
  in the \scp analysis because the bulk of the flux from the secondary
  component fell outside of the slitlet dimensions.  {\bf Top:}
  Objects 0336-0380 and 1643-1687.  {\bf Bottom:} Objects 2342-2525
  and 1057-2462.  The photometric close pair candidates are circled
  and labelled in the images.  Object 1057-2469 is two objects, the
  fainter of which is the photometric candidate.  As in
  Figure~\ref{Rband_2D}, each image is $\sim15''$ on a side and the
  labelled rectangle replicates the slitlet dimensions and orientation
  for that particular observation. The images were obtained using LRIS
  except object 2342-2525 which was obtained using COSMIC.  The four
  LBGs here, and two in Figure~\ref{phot_pairs}, display dominant \lya
  emission. }
\label{Rband_2Dnon}
\end{center}
\end{figure}

In addition to the five \scpp, we find six LBG close pair candidates
with separations $<15$ \kpcp.  These were not included as \scp by
definition because the bulk of the emission of the secondary galaxy
did not fall into the dimensions of the slitlet. Four of the six
SSCpair candidates are shown in Figure~\ref{Rband_2Dnon}, with the
remaining two shown in Figure~\ref{phot_pairs}.  Future spectroscopy
will determine whether the six candidates are indeed associated.

\begin{figure*}
\begin{center}
\scalebox{0.64}[0.66]{\rotatebox{90}{\includegraphics{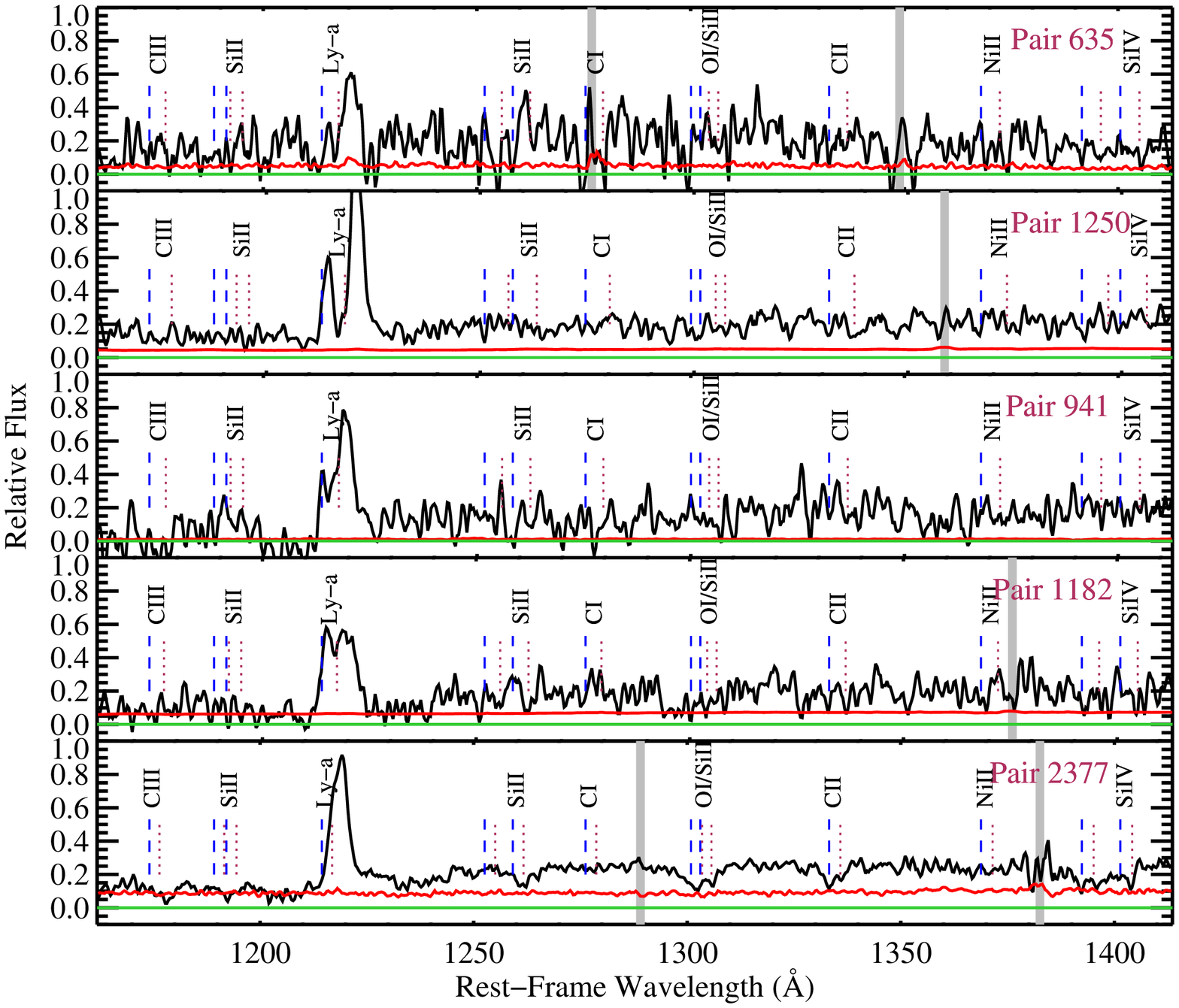}}}
\scalebox{0.15}[0.54]{\includegraphics{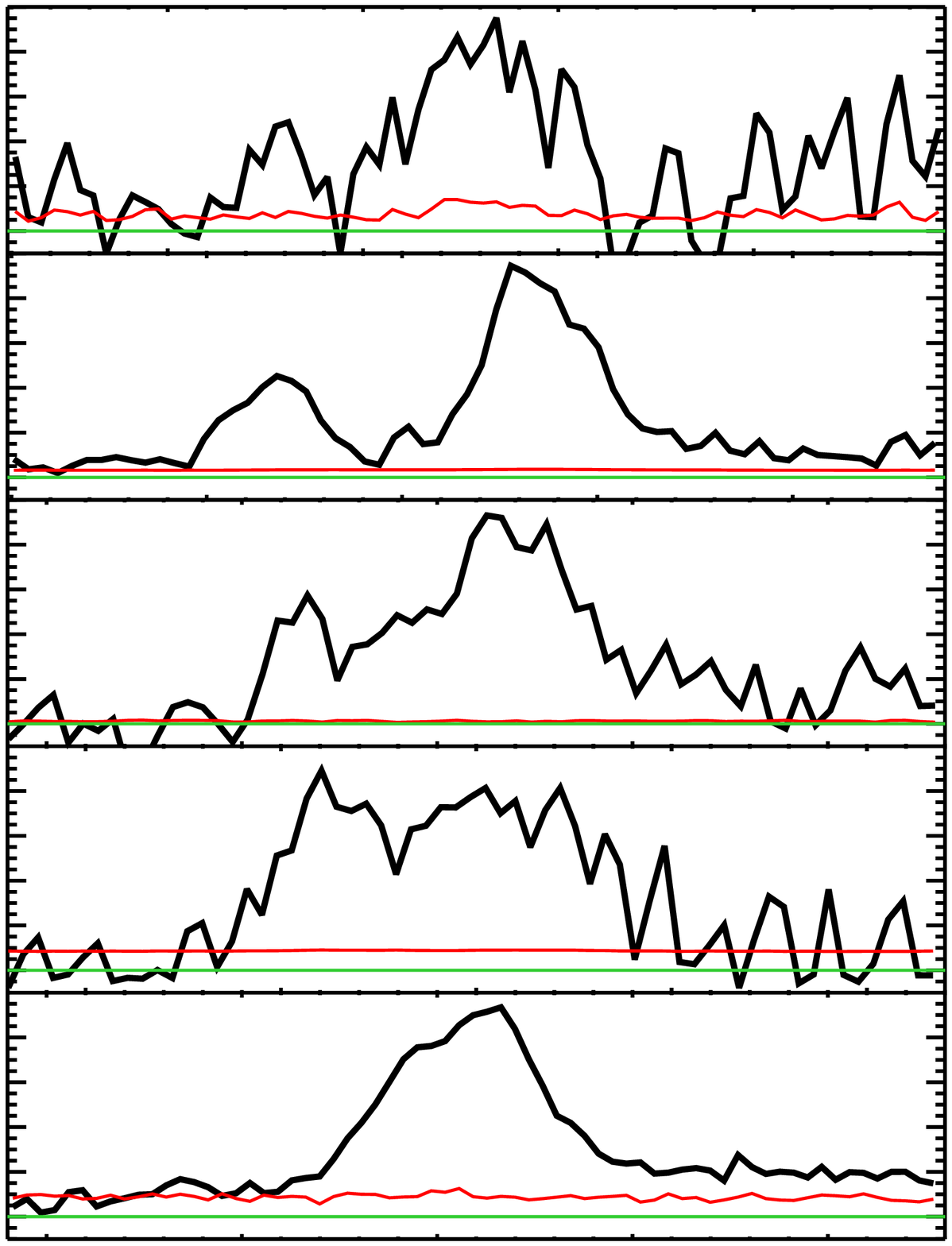}}
\caption
{\small {\bf Left:} Spectra of the five LBG serendipitous
  spectroscopic close pairs shifted to the rest-frame and box-car
  smoothed by 3 pixels.  As discussed in the text, these spectra are
  the merged 1-D spectra of the two very close or overlapping spectra.
  Although this aids in the identification and comparison of the
  systems, it blends and can dilute the significance of individual
  galaxy features.  Dashed (blue) and dotted (maroon) vertical lines
  mark the prominent ISM transitions of the apparent lower and higher
  redshift galaxy, respectively, and are labelled.  For clarity, not
  all lines used in the fit are shown.  The red (grey) line reflects
  the $1\sigma$ per pixel error array, the green (light grey)
  horizontal line indicates zero flux, and the thick solid (grey)
  vertical lines mark the positions of bright night sky emission
  lines.  {\bf Right:} 1-D spectral segments of the five pairs {\it
  (not smoothed)} that focus on a 24\AA~region (rest-frame) centred
  around the merged double-peak \lya emission features.}
\label{1Dpairs}
\end{center}
\end{figure*}

The small separations and overlap of the faint SSCpair spectra make
the clean extraction of individual components difficult.  As a result,
we present the merged spectra\footnote{Pair 0957-0941 has the largest
separation of the \scpp.  We made efforts to extract the individual
components of the low S/N spectra using conventional routines, but
found each spectrum was contaminated with a non-negligible amount of
flux from the other galaxy.  We present the merged spectra for this
pair in Figure~\ref{1Dpairs} for consistency.}  in
Figure~\ref{1Dpairs}, along with spectral segments focusing on the
\lya features.  Presented in this way, the two components of each
system can be directly compared, but have blended noise, continua, and
features that include absorption effects from the \lya forest blueward
of \lya emission.  Each component of the \scp shows \lya in emission
and we show in \S\ref{spectro} that this is not a selection bias.  The
presence of \lya emission from each component results in double-peak
\lya emission profiles for all five merged SSCpair spectra.

\begin{figure}
\begin{center}
\scalebox{0.40}[0.35]{\rotatebox{90}{\includegraphics{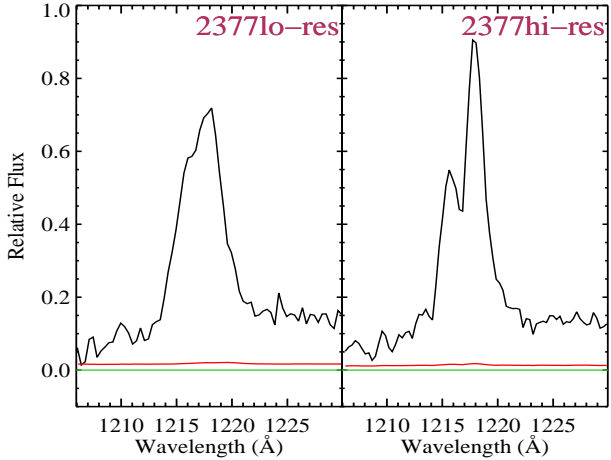}}}
\caption
{\small Spectra comparing the \lya emission for 1643-2377 obtained
  with two different resolutions.  {\it Left:} The merged
  low-resolution discovery spectrum ($\sim9$\AA~ resolution).  {\it
  Right:} The higher signal-to-noise higher-resolution
  ($\sim4$\AA~resolution) follow-up spectrum showing two distinct
  peaks.}
\label{2377}
\end{center}
\end{figure}

We remark that the double-peak \lya emission features result from
merging the spectra and differ from the double-peak \lya feature
predicted for a static or expanding shell of gas arising from a single
LBG \citep[e.g.,][]{tt99,m03,verhamme06,verhamme08}.  We describe
below that the \lya emission features of the \scp do not result from
single galactic-scale expanding shells as is evidenced by the two
spatially offset \lya emission peaks and continua in each SSCpair 2-D
spectrum, double morphology, and the direct correspondence between the
spatial offsets of the 2-D \lya features and continua and the flux
peaks in the images.  We note that an accurate measurement of the
small \lya emission velocity offset in the 1-D spectra of SSCpair
1643-2377 is difficult.  In that case, we used the high-S/N, higher
resolution spectroscopic data (Figure~\ref{2377}) to cleanly measure
the \lya peak offset and find that the measurement agrees with our
low-resolution value within the uncertainty.

Table~\ref{pair_vel} lists the relevant information of the five \scpp.
All reported velocity differences between the two components are
corrected for offsets in wavelength resulting from their spatial
separation as determined by the associated flux peaks in the images
(\S~\ref{spectro}).  We see that the two components have projected
physical separations of $\sim3-13$ \kpc in the images and 2-D spectra.
We remark that because these are projected separations, they
represent the minimum actual separations.

O and B stars are largely responsible for the rest-frame UV flux over
the wavelengths probed by the optical broadband filters used in this
survey.  As a result, we interpret the continua and emission lines
from the each SSCpair component as a major young star forming region
or a distinct LBG.  In the following sections, we discuss whether the
results favour an interpretation that the \scp are coalescing single
massive galaxies with two violent starbursts or an interpretation that
the \scp are two separate close and/or interacting galaxies.  The
lines of evidence presented below argue that these are separate
components that are inconsistent with the single galaxy
interpretation.  We use morphological and spectroscopic information
obtained from the \scpp, space-based investigations of local LBG
analogues, and LBG ground- and space-based observations and
relationships found in the literature to help interpret the \scp and
to help shed light on the rest-frame UV behaviour of LBGs.

\begin{table*}
\centering
\caption{\normalsize Serendipitous spectroscopic close pairs}
\label{pair_vel}
\footnotesize
\begin{tabular}{lcccccccccc}
\hline
Pair & R.A.$^a$ & DEC.$^a$ & m$_R$ & $z_{Ly\alpha}$$^b$ & 
$z_{ISM}$$^b$& $\Delta v_{Ly\alpha}$$^c$ & $\Delta v_{ISM}$$^c$ 
& Slit$^d$ & $\Delta r$$^e$ & Seeing$^f$\\
\hline
0056-0635 & 00:59:27.722 & +01:41:37.37 & 25.0, 25.2 &         
 3.367,3.386 & 3.363,3.374 & 870 & 350 & 1.5 & 9.5 & 0.8; 1.4\\
0336-1250 & 03:39:17.199 & --01:31:29.10 & 23.4$^g$& 
 2.824,2.845 & 2.819,2.835 & 960 & 575 & 1.5 & 8.2 &0.9-1.1; 0.6\\
0957-0941 & 09:57:48.414 & +33:11:28.43 & 24.8, 25.0 &
 2.680,2.688 & 2.678,2.685 & 480 & 400 & 1.5 & 13.3 & 0.9; 1.0\\
1013-1182 & 10:15:57.234 & +00:21:55.06 &24.6$^g$& 
 2.779,2.792 & 2.775,2.786 & 730 & 555 & 1.3 &  2.6 & 0.9; 0.9\\
1643-2377$^h$
          & 16:44:48.314 & +46:27:08.23 & 22.2$^g$ & 
 3.036,3.044 & 3.029,3.036 & 430 & 360 & 1.5 &  4.7 & 0.8; 0.7\\
\hline
\end{tabular}
\begin{tabular}{l}
$^a$ Coordinates of the LBG flux centroid used to mill the multi-object 
  spectroscopic slitlet.\\
$^b$ Assigned redshifts for the respective features of the two
  identified components\\
$^c$ Velocity offset in \kms between the respective features, corrected for 
  angular separation\\
$^d$ Width of multi-object spectroscopic slitlet in arcsec\\
$^e$ Projected angular component separation in the images in physical kpc\\
$^f$ Spectroscopic FWHM in arcsec derived from night sky 
  emission lines; R-band seeing FWHM in arcsec\\
$^g$ Integrated magnitude of both sources\\
$^h$ Values are from the low-resolution data for consistency
\end{tabular}
\end{table*}


\subsection{Morphology}\label{morph}

Space-based images of \zzz LBGs \citep[e.g.,][]{law07} show a wide
range of morphologies, such as concentrated single nucleated sources,
multiple nucleated sources, and diffuse structures.  Although complex,
typical LBGs have half-light radii of $\lesssim0.''3$
\citep{g00,ferg04}, or $\lesssim2$ \kpcp, and appear as near-point
sources in ground-based images.  We inspect the flux contours of the
\scp in the LRIS ground-based $u'$BVRI images\footnote{LBGs at \zzz
are colour-selected as $u'$-band ``drop-outs'' because the decrement
in their continua shortward of the Lyman limit ($912$\AA, rest-frame)
is redshifted into the $u'$-band.  Therefore \zzz LBGs are extremely
faint (R $\ge27$) or not detected in this bandpass.  Inspection of the
$u'$-band images reinforces the non-detection of both components
detected in the longer-wavelength bands and helps to eliminate
low-redshift line-of-sight interlopers.} for departures from an
unresolved source in the form of significant extended features,
elongations, or objects with more than one discernible flux peak.

The VRI broadband filters probe the star forming rest-frame continuum
of the \zzz LBG sample longward of \lyap.  We find that the centroids
of the flux peaks in the ground-based VRI images are accurate to
$\sim0.''4$ (approximately the seeing half-width half maximum) aided
by the near point-source profiles of LBGs.  Each LBG that meets the
2-D and 1-D criteria discussed below exhibits two measurably separated
flux peaks in the VRI images and have projected centroid separations
of $\sim3-13$ \kpcp.  The separations for \scp 1013-1182 and 1643-2377
are at or near the resolution limit.  The double nature of 1013-1182
relies largely on other measurements presented in this work such as
the two spatially offset \lya emission peaks in the 2-D spectra.  For
1643-2377, the high S/N from the intrinsic luminosity of this system
allows a secure centroid measurement.  Finally, we note that the
broadband continuum luminosity and \lya emission from each SSCpair
component falls in the range of typical m$_R\le25.5$ LBGs at \zzzp.
Consequently, each SSCpair component has LBG-equivalent estimated star
formation rates \citep[$\sim10^{1}-10^{3} M_\odot$
yr$^{-1}$,][]{pap01,aes01,aes03}.

At low to intermediate redshifts, the UV and optical studies of
compact UV-luminous galaxies \citep[UVLGs; e.g.,][hereafter
O08]{heckman05,grimes07,overzier08} have made great strides in our
understanding the observed LBG UV morphology at high redshift.  UVLGs
are in every practical way local analogues to LBGs having similar
mass, gas content, half-light radii, UV colours, metallicity, and star
formation rates.  O08 find that although the objects in their sample
do not show convincing evidence for mergers or interactions from the
HST rest-frame UV imaging alone, {\it every UVLG in their sample shows
evidence of interaction in the rest-frame optical images}.  They
conclude that mergers are the main mechanism for the observed star
formation in these systems and find that the bulk of the star
formation arises in several compact ($\sim100-300$ pc) regions.

O08 deconvolve the pixel scale and dim the surface brightness of their
$z\sim0.1$ UVLG sample to match the depth and resolution of \zzz LBG
observations in the GOODS, COSMOS, and UDF fields.  The multiple
bright star forming regions of the UVLGs blend into $1-3$ bright
clumps with separations of $\lesssim0.5''$ when redshifted to \zzz and
accurately reproduce the space-based morphology of LBGs.  For the
ground-based image resolution in our survey, these clumps would appear
as a single near-point source for all but the most separated cases,
which would be detected as \scp depending on component properties.
Perhaps the most convincing evidence of this is the case of SDSS
J080844.26+394852.4 (hereafter UVLG 0808) in the O08 sample.

We find that all UVLGs in the O08 sample, except for UVLG 0808, have
compact rest-frame UV morphologies and, if placed at \zzzp, would be
detected as single near-point sources\footnote{Object SDSS
J092600.41+442736.1 shows rest-frame H$\alpha$ emission with a
separation that is at the limit of what is discernible as two objects
in our ground-based survey.  It did not receive HST rest-frame UV
imaging, so it is unknown at this point whether it has two-component
UV flux that would be detectable as a close pair at \zzzp.  Strong
detection in the rest-frame UV is unlikely because one component is a
much redder galaxy that shows an old, or dusty, population in the SDSS
spectrum.}  in our survey.  UVLG 0808 has a fainter companion
$\sim2''$ to the SE that exhibits star formation flux in the UV image.
An equivalent separation at \zzz would be discernible in ground-based
images.  In addition, the companion has a compact strong detection in
the rest-frame H$\alpha$ image and is the only UVLG in the sample that
is detected as two separate galaxies by the Sloan Digital Sky Survey
(SDSS).

We search the SDSS spectra of the O08 sample for evidence of
double-peak emission in the optical nebular lines.  These lines are
not subject to outflows as are \lya emission lines (\S\ref{lya}) and
should closely trace the systemic velocities.  The multiple star
forming regions of seven of the eight systems are located within
$\sim1''$ and the component relative velocities are near or below the
spectral resolution of SDSS, thus these UVLGs show single-peak
emission.  However, UVLG 0808 shows evidence of a double-peak feature
as witnessed in the Balmer lines.  Because the two components of UVLG
0808 have an $\sim2''$ separation, a portion of rest-frame optical
flux from the companion galaxy is expected to fall into the SDSS fibre
and result in the weaker offset emission line peaks ($\Delta v
\sim280$ \kmsp).  UVLG 0808 is the only object in the sample that (1)
has a well-separated companion, (2) appears as two star forming point
source galaxies in the rest-frame UV, and (3) shows evidence of
double-peak emission in the 1-D spectroscopy.  Observed at \zzzp, a
system with similar properties would appear by our definition as an
SSCpair.

Projected component separations of $\gtrsim4$ \kpc are necessary for
individual identification of UVLGs in the O08 sample and LBGs in our
survey at the resolution of SDSS and LRIS, respectively.  The
remarkable similarities between the properties of UVLGs at low to
intermediate redshift and LBGs at high redshift, and the evidence that
interactions provide the bulk of the star formation in UVLGs, leads us
to conclude that the separations and emission profiles observed in our
\scp likely result from two interacting LBGs.  We investigate this
further in the following sections and discuss the spectroscopic and
morphological properties of LBGs found in the literature.

\subsection{Spectroscopy}\label{spectro}

The 2-D spectra are individually examined for signs of multiple
spectra beyond the usual pipeline reduction of the targeted objects
and distinct serendipitous objects.  We search for evidence of closely
spaced or overlapping spectra, and/or multiple \lya emission or
absorption features.  Alongside the image and 2-D analysis, we inspect
the 1-D spectra for evidence of more then one system from redshift
fits to $\sim20$ expected interstellar features.  The actual number of
features fit for a given LBG depends on the redshift and wavelength
range probed by the MOS slitlet and does not include lines that fall
near bright night sky emission features which can be difficult to
extract cleanly from the faint spectra.  The spectra in our sample
have continua with relatively low S/N ($\sim3-10$) and are typical of
LBG observations at $z\sim3$.

\lya is the most prominent rest-frame UV feature in \zzz LBGs and, in
contrast, can be detected at a high significance.  The LBG spectra
compiled to date show roughly 50\% of all LBGs dominated by \lya in
emission, with the remaining dominated by \lya in absorption
\citep[][and this sample]{aes03}.  Net EW values range from strong
emission, $>100$\AA~EW, to damped absorption (column densities of
N(H\textsc{i}) $\ge 10^{20.3}$ atoms cm$^{-2}$).  Here we term LBGs
that are dominated by \lya emission in their spectra, eLBGs, and those
dominated by \lya in absorption, aLBGs.  After a thorough search for
\scp of all types, we observe that each SSCpair component, with no
exception, displays \lya in emission.  Thus, in our nomenclature, all
SSCpair components are eLBGs.

The \lya emission features and continua in the 2-D spectra for every
SSCpair have the same separations in the spatial direction as the
image centroid separation along that same calculated direction.  The
one-to-one correspondence helps to confirm the distinct two-component
nature of the \scp and enables \lya emission and star-forming continua
assignment to each peak in the images.  We use the image centroids
when correcting the \lya and ISM absorption-line velocity offsets for
their angular separation.  The separations in the dispersion direction
are complicated by the fact that \lya emission has a large range of
velocity offsets with respect to the systemic redshift (see
\S~\ref{lya}).  Nevertheless, in the cases where the flux peaks in the
images have small separations in the spatial direction, large
differences between the offsets of the two \lya features can provide
additional evidence for two major star-forming components.

In practice, detecting double \lya emission can prove to be easier
than double \lya absorption when component velocity offsets and
angular separations are small.  Thus, it is possible, that even after
a thorough search, we are missing a fraction of LBG spectroscopic
close pairs that only exhibit \lya in absorption because of the low
S/N of the spectra and broad nature of the \lya absorption features.
However, LBGs displaying \lya in absorption typically have stronger
ISM transitions that can facilitate the detection of a velocity offset
between two overlapping spectra \citep[][C05]{aes03}.  We find no
aLBGs at larger radii where both the angular separation and velocity
offsets would resolve the \lya features of the two systems and no
clear evidence for double sets of ISM absorption transitions redward
of \lya in aLBG spectra.  In addition, we do not find \scp having one
component in emission and one in absorption in either the 1-D or 2-D
spectra.  Finally, all spectroscopic components of the SSCpair
candidates (Figure~\ref{phot_pairs} \&~\ref{Rband_2Dnon}) show
dominant \lya emission, with none exhibiting dominant \lya absorption.
These findings are discussed further in \S\ref{ecp}.  As a result, the
observation that all \scp are eLBGs is not primarily a selection
effect.

\subsubsection{\lya offsets}\label{lya}

We measure corrected \lya velocity offsets for the \scp of $\Delta
v_{Ly\alpha}\sim400-1000$ \kmsp.  To expand on the difficulties in
rest-frame UV continuum interpretation mentioned earlier, \lya
emission in LBGs is observed to be redshifted by $\sim450\pm300$ \kms
with respect to the systemic redshift \citep[C05][]{a03,aes03}.  The
velocity offsets likely result from galactic-scale outflows caused by
stellar and supernova-driven winds \citep[e.g.,][]{heckman02,a03}.
\lya photons are absorbed by the approaching outflow while \lya
photons scattered off-resonance from the receding shell can escape.
This geometry results in \lya emission peak velocity offsets that are
redshifted with respect to the systemic redshifts and dependent on the
host galaxy properties. Consequently, the observed difference between
the peaks of the \lya features cannot be used directly as a measure of
the systemic velocity offsets of the SSCpair components.

However, under the following assumption, we can investigate the nature
of the component \lya offsets using the \lya offset distribution of
the LBG population.  Because of the one-to-one correspondence between
the $\sim3-13$ \kpc projected separations of the flux peaks in the
images and the \lya peaks in the 2-D spectra, we attribute the
double-peak \lya features to two distinct galactic-scale outflows.
Admittedly, this is a simplified picture but the relatively coarse
resolution of our ground-based observations blends the velocities of
any localised regions and can only probe net global outflows.

\citet{a03} find a velocity offset distribution between \lya emission
and the ISM absorption lines ($\Delta v_{Ly\alpha-ISM}$) of $\Delta
v_{Ly\alpha-ISM}=614\pm316$ \kmsp.  Because LBG ISM absorption lines
are blueshifted on average with respect to the systemic redshift by
$\sim-150\pm160$ \kms \citep[][and this survey]{a03,aes03}, this
results in the above-mentioned $450\pm300$ \kms \lya velocity offset
distribution.  In addition, \citet{aes03} find that LBGs with dominant
\lya emission (Group 3 and Group 4 in that sample, or eLBGs in our
nomenclature) show smaller values of $\Delta v_{Ly\alpha-ISM}$ and
therefore evidence of weaker outflows.  If we assume that eLBGs have a
similar velocity offset distribution in behaviour and form to that of
the full LBG population, the eLBG-only distribution becomes $\Delta
v_{Ly\alpha-ISM}=518\pm266$ \kmsp, with eLBG \lya offsets from the
systemic redshift of $406\pm262$ \kmsp.

If the \lya emission in the \scp reflects outflow from two star
forming regions in single coalescing systems, we would expect the
velocity difference between the two \lya features, $\Delta
v_{Ly\alpha}$, to randomly sample either of the above \lya velocity
offset distributions.  A random sampling of two values using the
distribution of the eLBGs and full LBG population results in expected
differences of $\Delta v_{Ly\alpha}= 295\pm220$ \kms and $\Delta
v_{Ly\alpha}= \sim340\pm255$ \kmsp, respectively.  However, we find a
much larger $\Delta v_{Ly\alpha}=694\pm234$ \kms for the \scpp.  If we
add a line-of-sight velocity component originating from two regions in
a bound, massive LBG system by pulling from a $\lesssim250$ \kms
distribution of relative velocities
\citep[e.g.,][]{erb03,erb06,genzel08,law09}, we find an expected net
two-component velocity offset distribution of $\Delta
v_{Ly\alpha}\sim320\pm230$ \kms (or $\sim360\pm260$ \kms for the full
LBG population), which falls short of describing our data.

The correspondence between the 2-D \lya emission and continua and the
two flux peaks in the images forces us to conclude that the \lya
emission features of the \scp arise from either (1) single galaxies,
but do not randomly sample the behaviour observed in the full or
subset LBG population and that larger than typical $\Delta
v_{Ly\alpha}$ values are correlated with star forming regions having
substantial line-of-sight velocity contributions or (2) star forming
regions with velocity offsets too large to be single systems and are
truly pairs of LBGs.  In either scenario, the double-peak \lya
emission of the \scp lends insight into the behaviour of \zzz systems.

\subsubsection{ISM absorption}

Spectra showing two distinct sets of ISM absorption lines having large
velocity offsets together with separate 2-D spatial offsets provide
strong evidence for two independent systems.  We cross-correlate and
visually inspect the \scp for $\sim$20 strong ISM UV transitions that
include Ly-$\alpha$, Ly-$\beta$, and strong ISM features such as
Si\textsc{ii}$\lambda1260$,
O\textsc{i}/Si\textsc{ii}$\lambda\lambda1302, 1304$,
C\textsc{ii}$\lambda1334$, Si\textsc{iv}$\lambda\lambda1393, 1402$,
Si\textsc{ii}$\lambda1526$, and C\textsc{iv}$\lambda\lambda1548, 1550$
\AA.  We find evidence for two systemic redshifts from both the low-
and high-ion transitions in the merged 1-D spectra.  The ISM velocity
offsets for the \scp are listed in Table~\ref{pair_vel}.  We caution
that the low S/N ($\sim3-10$) of the continua and the potentially
complex nature of the rest-frame UV behaviour of the ISM, make these
the least confident measurements in terms of significance when
compared to the other analyses presented here.  However, the ISM
measurements of the higher S/N low-resolution spectra of 0336-1250
($\sim6\sigma$) and 1643-2377 ($\sim11\sigma$) are more secure.

Our data provide tests of the validity of these low S/N measurements.
Object 0336-1250 was observed with three separate slitmasks for a
total integration time of 11700s and the luminous 1643-2377 was
observed with three exposures of 1500s each.  In an effort to mimic
the lower S/N of the other SSCpair total observations, we performed
fits to each of the three 0336-1250 slitmask observations and the
individual exposures of 1643-2377 separately.  We found agreement to
$\Delta z\le 0.0015$ between the fit to the individual exposures to
their respective combined spectrum.

The \scp show a corrected velocity difference between the two
identified sets of ISM lines of $\Delta v_{ISM}=448\pm109$ \kmsp.  As
mentioned above, ISM lines exhibit an $\sim150\pm160$ \kms range in
blueshift from systemic.  If we assume that the two sets of ISM lines
randomly sample this distribution, we would expect $\Delta
v_{ISM}=180\pm135$ \kmsp.  If we assume the \scp were single systems
with the bulk of the ISM absorption seen in the sightlines to two
well-separated star forming regions with bound relative velocities
$\lesssim250$ \kmsp, we would expect the combined effect to result in
$\Delta v_{ISM}\sim194\pm145$ \kmsp, which is much smaller than the
observed value.  As a result, the evidence in the ISM lines argue for
the \scp to be separate systems.

\subsubsection{Ly-$\alpha$ - ISM velocity difference}\label{lya-ism}

A comparison of the difference between the SSCpair \lya and ISM
features with the relationships found for LBGs is a further test of
the viability of the 1-D spectral features to help confirm the
double-galaxy nature of the \scpp.  As discussed above, \citet{aes03}
determined $\Delta v_{Ly\alpha-ISM}=614\pm316$ \kms for all LBGs and
$\Delta v_{Ly\alpha-ISM}=518\pm266$ \kms for eLBGs that are similar to
the \scpp.  Because we see two \lya emission peaks and evidence for
two sets of ISM absorption lines, we treat the SSCpair components as
individual systems and measure the $\Delta v_{Ly\alpha-ISM}$ for each
component separately using no angular separation correction.  This
results in a $\Delta v_{Ly\alpha-ISM}=475\pm205$ \kms when assigning
the highest redshift \lya feature to the highest redshift set of ISM
features and the lowest redshift \lya feature to the lowest redshift
ISM features.  A reassignment of the \lya and ISM features results in
the same central value but a larger scatter.  The $\Delta
v_{Ly\alpha-ISM}$ distribution found for the \scp agrees with that of
the LBG population.  Moreover, the distribution follows the trend
found in \citet{aes03} for LBGs displaying dominant \lya emission.

In treating the \scp as close pairs throughout this section, we would
expect the 1-D spectral features of each galaxy to reflect the
relationships and trends of the LBG population.  We find that the
$\Delta v_{Ly\alpha}$, $\Delta v_{ISM}$, and $\Delta v_{Ly\alpha-ISM}$
are in good agreement with this picture and are at odds with an
interpretation that the \scp represent starforming regions in single
systems.

\subsection{Comparison to previous related work}\label{previous}

We look for observational precedent in the literature to help
interpret the \scp by first exploring the LBG sample of \citet{aes06}.
This sample consists of deep spectroscopy of 14 LBGs at \zzz gathered
for the purpose of measuring the fraction of escaping continuum flux
shortward of the Lyman limit.  The observations are valuable for an
assessment here because they were selected in the same manner, lay
within the same redshift path, and have spectroscopy obtained using
the same instrument (LRIS).  Moreover, the observations are beneficial
here because they have a slightly higher resolution over the bulk of
the more relevant wavelengths ($\lesssim1400$\AA~rest-frame) and have
$\sim10\times$ longer exposure times for S/N of $\sim5-25$.  In
addition, the associated imaging data were obtained with similar or,
for the subset of space-based data, superior depth and resolution.

Of the 14 LBGs in that sample, nine are eLBGs.  The \lya EW
distribution of the sample is very similar to that of our full
spectroscopic LBG sample (K-S test, $p=0.6$).  One object, D3, is
considered a single system by \citet{aes06} but consists of two
galaxies (and separate spectra) with an angular separation of $1.''9$
($\sim15$ \kpc).  Three of the nine eLBGs are reported to show
double-peak \lya emission.  Because both components of D3 show \lya
emission, merging the two spectra would create a double-peak spectrum,
making it a fourth double-peak system (similar to 0957-0941 in our
sample).  Relevant to later discussion, of the two LBGs shown by
\citet{aes06} to exhibit measurable Lyman continuum flux, one is the
double-peak \lya emission system C49 and the other is the pair D3.
Narrow-band imaging of D3 shows that it is detected as a \lya ``blob''
in the sample of \citet{matsuda04} with \lya emission extending over
$17$ arcsec$^{2}$.

We now analyse the double-peak systems of \citet{aes03} as we did for
the \scp using the published information.  Looking at the velocity
offsets of the \lya feature of the four double-peak systems, we find
$\Delta v_{Ly\alpha}= 533\pm170$ \kmsp, cf. $\Delta v_{Ly\alpha}=
694\pm234$ \kms for our sample.  Shapley and coworkers measure one ISM
absorption redshift for each of the three \lya double-peak systems
(pair D3 has separate values).  If we assign the reported ISM
absorption redshifts to the highest redshift \lya peak, we find
$\Delta v_{Ly\alpha-ISM}=1165\pm667$ \kmsp.  This places these systems
at the high-end tail of the $\Delta v_{Ly\alpha-ISM}$ distribution for
the full LBG population, with one system having the highest ($1823$
\kmsp) velocity difference of the entire \zzz LBG spectroscopic
sample.  Assigning the features in this manner implies that it is
highly improbable that the double-peak systems are single LBGs.

Instead, if we assign the ISM absorption redshifts to the lowest
redshift \lya peaks we find $\Delta v_{Ly\alpha-ISM}=579\pm682$ \kmsp.
While the central value matches the expectations of eLBGs and is in
good agreement with LBGs in general, the large scatter of the three
systems is unrepresentative of the $\pm275$ \kms scatter that we find
for three-galaxy random samples taken from the LBG and eLBG data.  In
addition, we see similar relative offset and scatter behaviour when we
assign the higher redshift ISM features to the lower redshift \lya
emission (and vice-versa) in our SSCpair sample
(section~\ref{lya-ism}).  When inspecting the double-peak system C32,
we find that the lowest redshift \lya peak is blueshifted with respect
to the ISM absorption lines by $-183$ \kmsp.  If we assign only this
system to the higher \lya peak, then the three systems exhibit $\Delta
v_{Ly\alpha-ISM}=803\pm322$ \kms and the scatter is reduced to the
level of the expectations.  When including the identified pair D3
($\Delta v_{Ly\alpha-ISM}=579$ \kmsp) in the \lya double-peak sample,
we find $\Delta v_{Ly\alpha-ISM}=713\pm266$ \kmsp, which is more
consistent with the distribution and scatter of the eLBG and LBG
population.  Although matching ISM features to one of the \lya
emission peaks can make the double-peak systems fall in line with
typical LBGs, what should we make of the second \lya feature?
 
The above analysis shows that it is more likely that the three \lya
double-peak LBGs are close pairs similar to D3 and the SSCpair sample
as opposed to single systems.  Such an interpretation produces a
$\Delta v_{Ly\alpha-ISM}$ distribution in better agreement with the
expectations of the LBG population and significantly reduces the
scatter.  It is possible that the presence of a second set of ISM
absorption lines in the \citet{aes06} sample may have been unexpected
and ignored, identified and interpreted as strong ISM mixing, or
identified and considered of too low significance as a result of
blending.  Although the S/N of the \cite{aes03} LBG sample is higher
than what is typical for \zzz LBG spectra, it does remain relatively
low in that ISM absorption lines remain difficult to study with great
confidence.  Future investigation of the data (for example, by probing
nebular lines in the IR) could resolve this matter.

We remark that \cite{aes03} find $33$\% ($44$\% depending on slit
orientation for D3) of the eLBGs in their sample, and $21$\% ($29$\%)
of the entire sample, to be double-peak systems that meet our SSCpair
criteria.  The data imply that higher S/N and/or higher resolution
should reveal a larger fraction of \scp than that seen serendipitously
in our low-S/N, low-resolution survey, with the true fraction
potentially even higher.

Next, we examine the study of \citet{law07} that consists of
space-based imaging of 216 $z\sim2-3$ LBGs and ground-based
spectroscopy using LRIS with a similar configuration as our survey.
In that work, the morphology of LBGs is quantified using a
non-parametric analysis within $1.5''$ radius ($\sim13$ kpc) which
directly reflects the regime of the \scpp.  In addition, this is the
regime where systems are most sensitive to interaction that can
visibly change their morphology and star formation properties.  Law
and coworkers investigate the rest-frame UV morphology of their LBG
sample by quantifying the Gini parameter, colour dispersion,
multiplicity, and size.  The Gini parameter is a measure of the
nebulosity or nucleated nature of the system, the colour dispersion is
measured using V $-z$ broadband colours and corresponds to rest-frame
$\sim1500-2100$\AA~at \zzzp, and the multiplicity parameter quantifies
the multiple component nature of the sources.

\citet{law07} investigate the relationships between the morphological
parameters and the spectroscopic features \lya EW, low- and
high-ionisation ISM EW, and $\Delta_{Ly-\alpha - ISM}$.  As noted in
that work, only \lya EW shows a significant and consistent trend at
\zzz with all four morphological parameters.  The \lya relationships
indicate greater \lya emission EW with decreasing system size,
decreasing nebulosity (increasing Gini parameter), decreasing number
of nuclei (multiplicity) or nuclei separation, and bluer colours.  In
addition, and in reference to low-redshift UVLGs, a visual inspection
of the space-based images shows that systems with bright nuclei
typically have one or two components whereas systems with faint nuclei
typically have three or more components and are found in more diffuse
systems.

\lya EW is seen to be strongest for large values of the Gini parameter
indicating that nucleated sources are more likely to be eLBGs and
diffuse sources are more likely to be aLBGs.  The \lya EW increases
monotonically over the multiplicity range with values that indicate
two strong double nucleated sources (multiplicity parameter value
$\sim10$) down to a single nucleated source (multiplicity parameter
value $0$).  Finally, the \lya EW increases as the size of the systems
decrease and the colour becomes bluer.  \citet{law07} show that each
of these trends at \zzz has a $\chi^2$ significance of $\ge98$\%.
Taken together, these data can be interpreted to show that double
nucleated LBGs exhibit \lya emission and that the \lya emission EW
increases as the components become bluer and as the separation between
the two components decreases, or in other words, as the overall object
size becomes progressively smaller for a given multiplicity.

The SSCpair observations are consistent with this behaviour and the
interpretation that the \scp are interacting bright double nuclei, or
two single nuclei LBGs.  The double nucleated sources with larger
separations are those discernible as \scp in ground-based data, as
also indicated by the work of O08.  Diffuse systems, and diffuse
systems with multiple fainter clumps, would be observed as single LBGs
in ground-based data.  Moreover, such systems are LBGs with dominant
\lya absorption from the Law et al. relationships.  Strong \lya EW and
blue continua can be evidence for recent star formation and/or
decrease in dust and gas obscuration.  The \lya EW of the \scp in
order of decreasing projected separation is $26, 31, 102, 56,$ and
$91$\AA.  Although the \scp sample is too small to determine a trend,
the \scp have similar separations as the double nuclei LBGs in the
\citet{law07} sample, have blue colours, and exhibit \lya emission
with EWs that are in the upper quartile of the full LBG sample.

We comment that the work of \citet{law07} does not investigate the
behaviour of LBGs with separations larger than $\sim13$ kpc.  We probe
this regime in \S\ref{close_pairs} and address statistics applicable
to this discussion in \S\ref{ecp}.  In the next section, we test the
expectations of finding the \scp by measuring the distribution of LBG
close pair separations in our full survey and a larger survey from the
literature and by testing LCDM predictions using a high-resolution
cosmological simulation.  We show that the number of \scp are indeed
expected and, as a result, the investigation of this and other similar
samples offers a unique window into the behaviour and properties of
close/interacting high redshift galaxies.


\section{SERENDIPITOUS PAIR EXPECTATIONS FROM CLOSE PAIR SEPARATIONS}
\label{close_pairs}

To investigate the distribution of LBGs on small scales ($<500$
\kpcp), we measure the number of LBG pairs versus separation from
observation.  To help test the universality of the results and to
improve the statistics, we augment an investigation of our survey with
the photometric and spatial information from the larger survey of
\citet{s03}, hereafter S03, consisting of $\sim2300$ colour-selected
and $\sim800$ spectroscopically identified LBGs in 17 separate fields.

\subsection{Pair definition}\label{define}

For the C05 survey, we consider only LBGs that are identified in our
images by {\it SExtractor} with m$_R\le 25.5$.  The LBGs were selected
using $u'$BVRI colour criteria

\begin{equation}
(u'-B)_{AB} > 1.1
\end{equation}
\begin{equation}
0.6 < (B-R)_{AB} < 2.1
\end{equation}
\begin{equation}
0.6 < (B-I)_{AB} < 2.1
\end{equation}
\begin{equation}
(u'-V)_{AB} > 1.6
\end{equation}
\begin{equation}
(V-R)_{AB} < 0.6
\end{equation}
\begin{equation}
(V-I)_{AB} < 0.6.
\end{equation}

In order to provide more reliable photometric statistics, we use only
colour-selected LBGs that meet the full criteria in the seven of the
nine fields that have high-quality five filter imaging and LBGs that
meet four-filter criteria in the remaining two fields (see C05 for
details regarding the $u'$BVRI selection criteria and the specifics of
the color criteria for each of the nine fields).  For the S03 survey,
we use all LBGs that meet the $(G-$$\cal{R}$$)\le 1.2, (U_n-G)\ge
(G-$$\cal{R}$$)+1.0$ used in that work.  These color-selection
criteria are tested using $\gtrsim300$ and $\gtrsim1300$ spectra in
the respective samples.

\begin{figure}
\begin{center}
\scalebox{0.35}[0.33]{\rotatebox{90}{\includegraphics{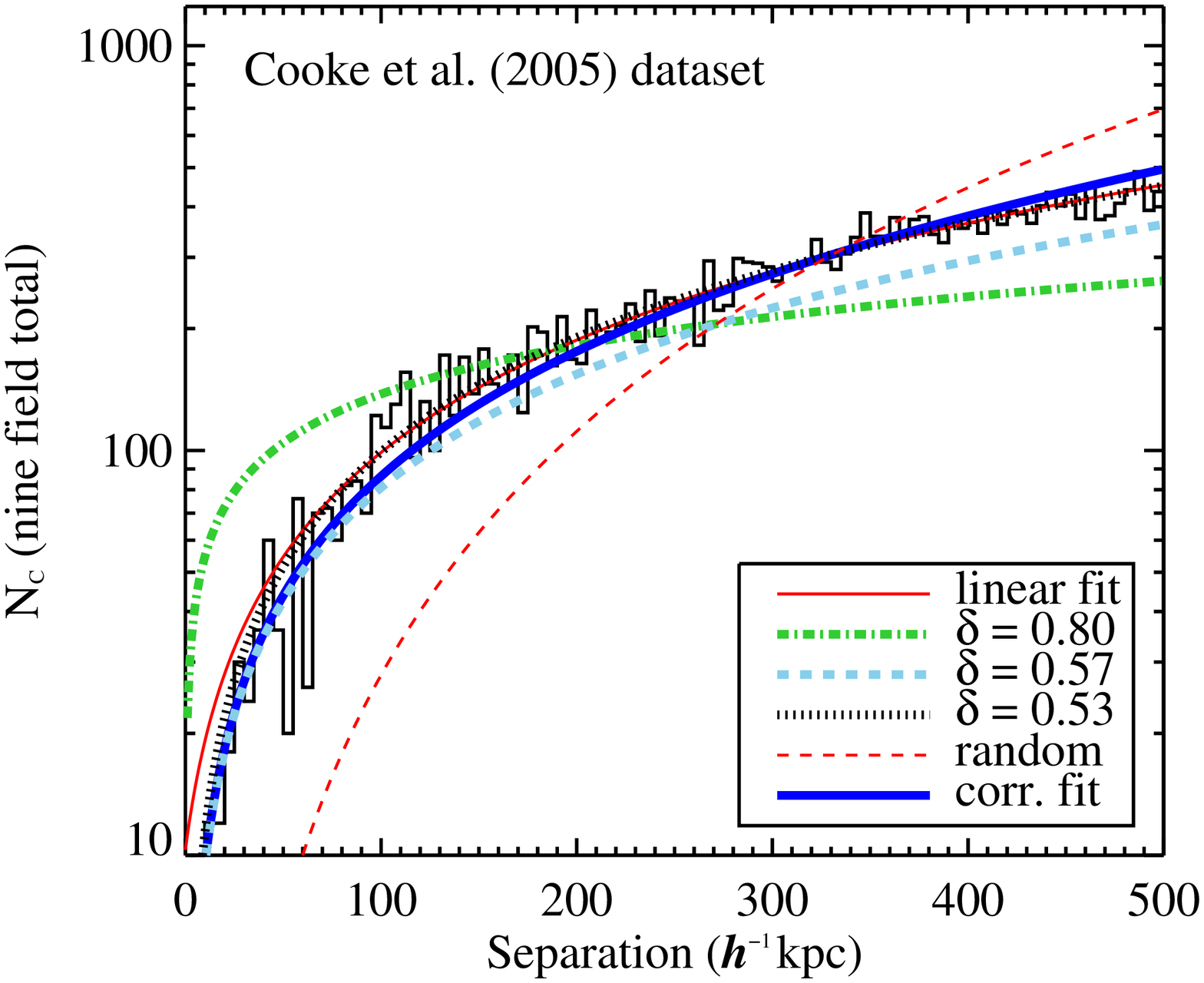}}}
\scalebox{0.35}[0.33]{\rotatebox{90}{\includegraphics{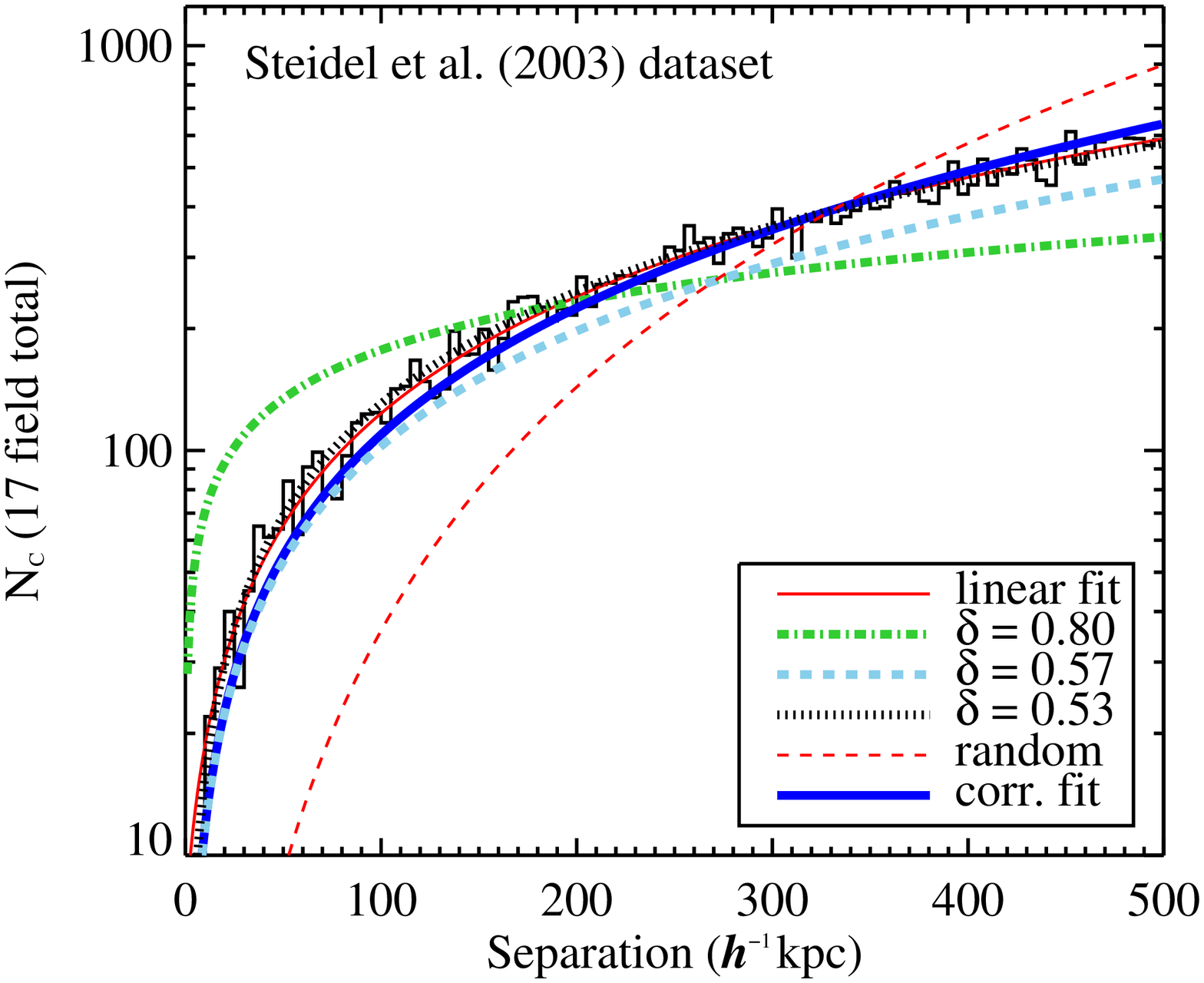}}}
\scalebox{0.35}[0.33]{\rotatebox{90}{\includegraphics{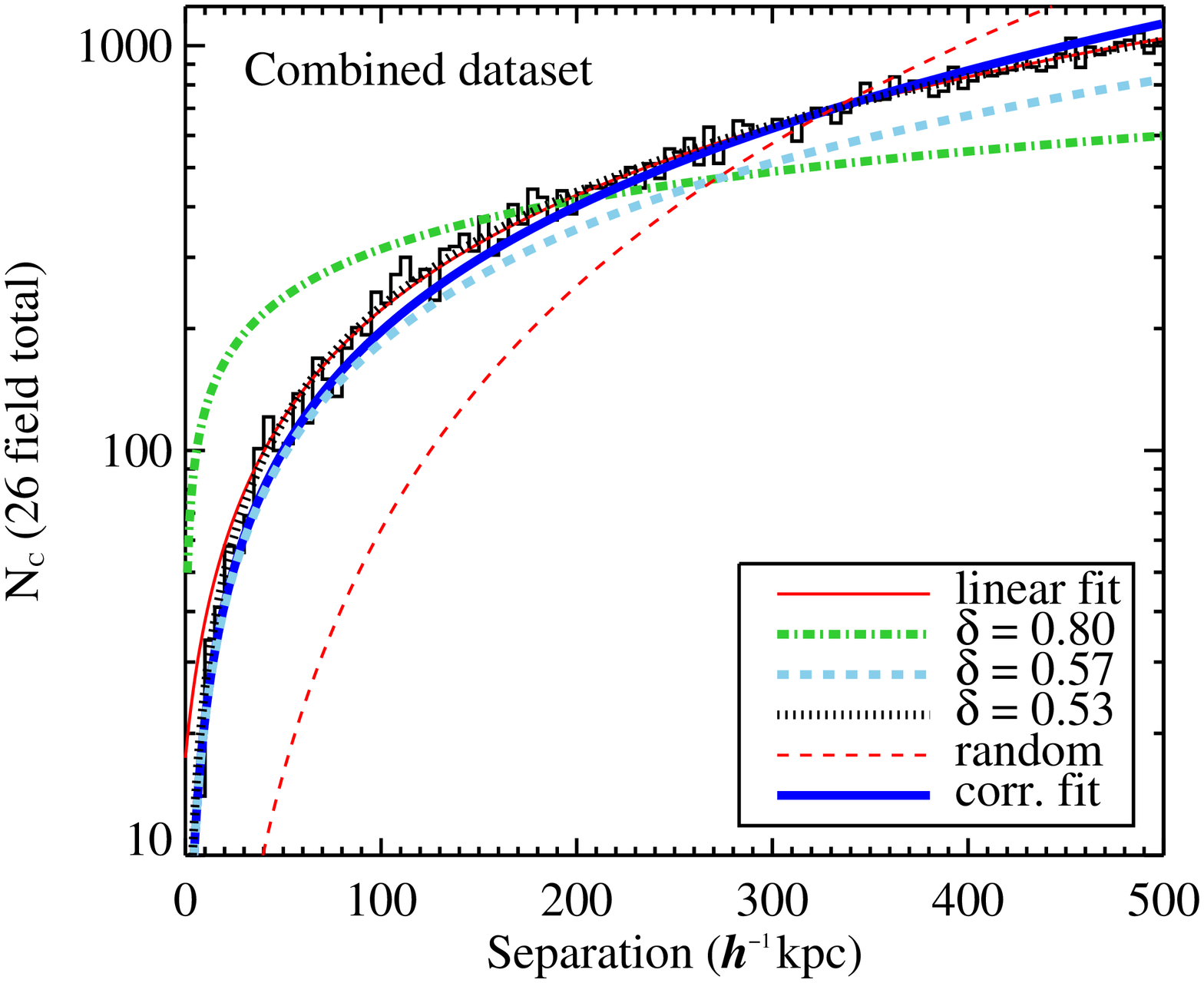}}}
\caption
{\small Histograms of pair separations versus radius.  {\bf Top:}
  Number of LBG pairs in our survey binned by $5$ \kpcp.  Overlaid,
  and more distinguishable on small scales in
  Figure~\ref{pairhist_50}, is a linear fit (thin solid red line), a
  power-law fit (thick dotted black line), and the corrected
  two-point angular correlation function of the form $\omega (\theta)
  = A\theta^{-0.57}$ (thick solid blue line) fit to the data (see
  text).  The contribution from LBG pairs to the angular correlation
  function fit is shown by the thick dashed (light blue) line, with
  the random interloper pairs comprising the remaining contribution.
  The thin (red) dashed line indicates the form of the random pair
  contribution normalised to data.  As a comparison, the LBG
  contribution for an angular correlation function with slope
  $\delta=0.8$ (thick dot-dashed green line) is shown.  {\bf Centre
  and bottom:} The same as the top panel, but for the survey of
  \citet{s03} and the combined dataset from both surveys,
  respectively.}
\label{pairhist_500}
\end{center}
\end{figure}

LBG colour selection at \zzz is very efficient in filtering out
objects not in the redshift path of interest (interlopers).  As
determined from the two spectroscopic samples, the fraction of
interlopers in each survey is similar ($\sim0.20$) when considering
the particular treatment of the unidentified low S/N spectra.
Differences in the photometric selection functions and selection
efficiencies between the two surveys are small and are quantified and
corrected for using the information from the spectroscopic samples.
Figures~\ref{pairhist_500} \&~\ref{pairhist_50} present the close
pairs versus separation for this survey, the survey of S03, and the
combined dataset.  The two figures emphasise large and small scales,
respectively.  The fall-off in the number of pairs at radii
$\lesssim10$ \kpc (best seen in Figure~\ref{pairhist_50}) results from
the resolution of the images and the ability for {\it SExtractor} to
discern close galaxies as separate.

\subsection{Pair separations}\label{pairsep}

\begin{figure}
\begin{center}
\scalebox{0.35}[0.33]{\rotatebox{90}{\includegraphics{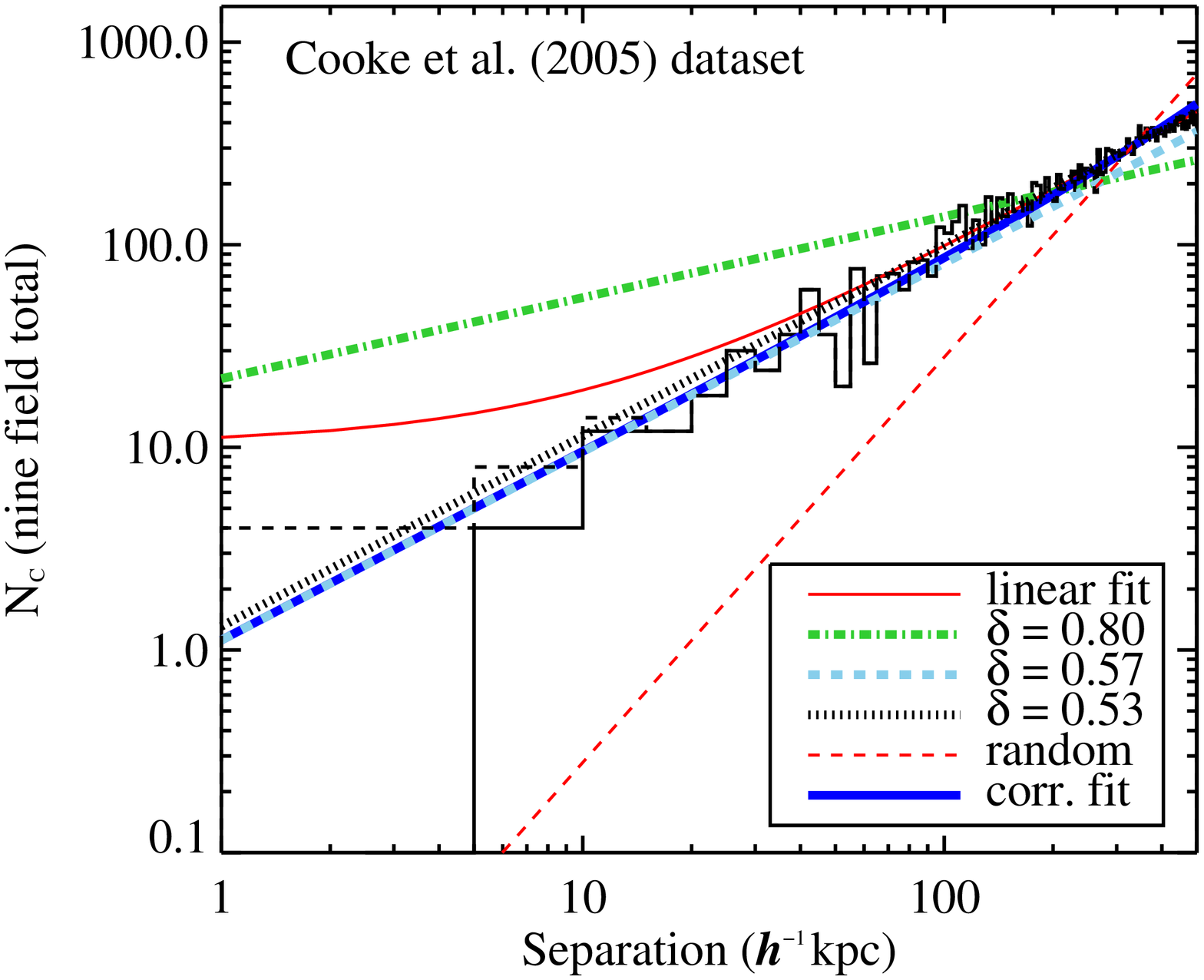}}}
\scalebox{0.35}[0.33]{\rotatebox{90}{\includegraphics{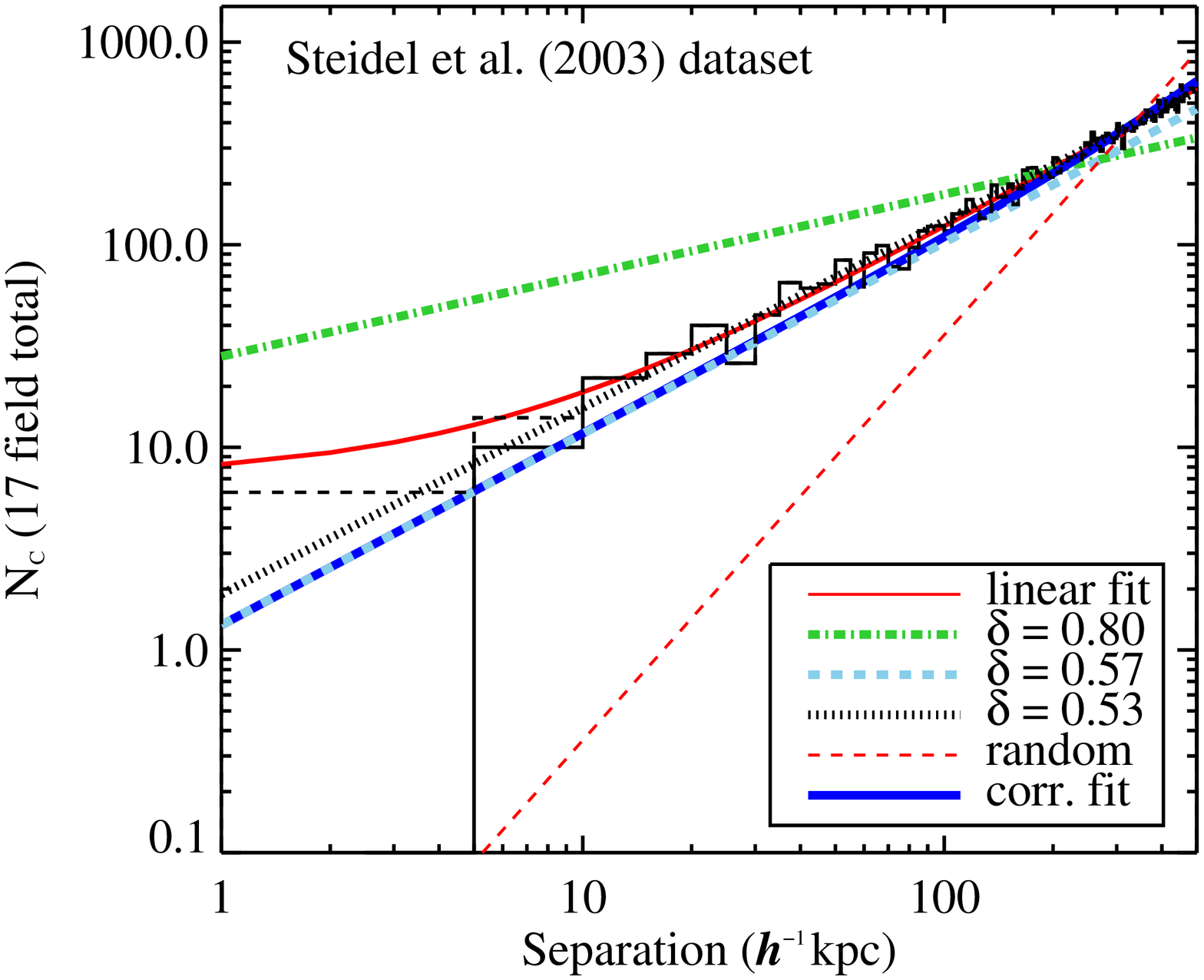}}}
\scalebox{0.35}[0.33]{\rotatebox{90}{\includegraphics{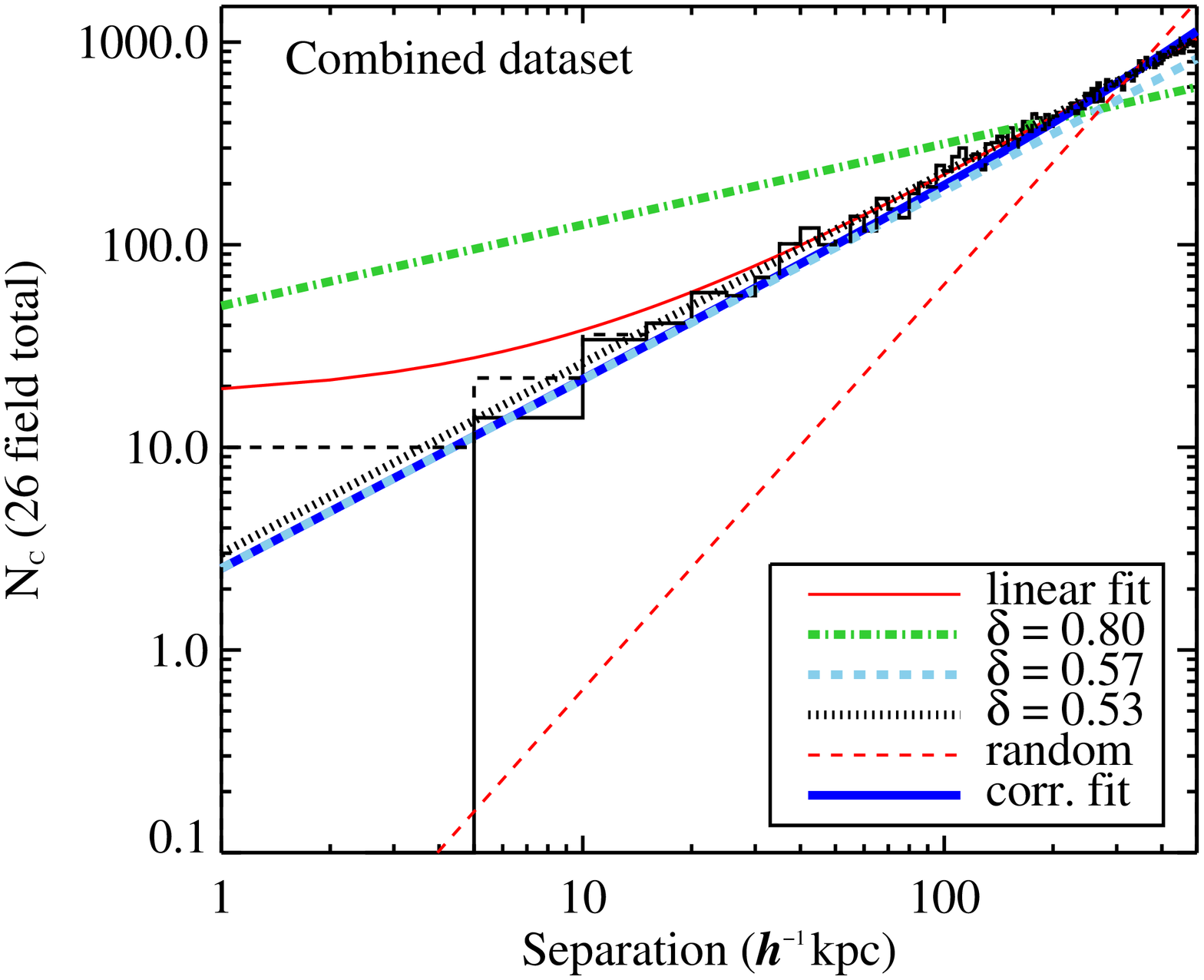}}}
\caption
{\small Same as Figure~\ref{pairhist_500} but in log/log space to
  better discern the behaviour on small ($\lesssim20$ \kpcp) scales.
  The photometric close pairs (solid histogram) fail at separations
  smaller than $\sim7$ \kpc because of the resolution of the images
  (also see text).  The number of close pairs with the addition of the
  \scp is represented by the dashed histogram.  The top panel shows
  the close pairs and \scp presented in this work.  The central panel
  shows the close pairs for the survey of \citet{s03} and includes
  predicted \scp using an expected distribution based on the observed
  distribution found in our survey.  The bottom panel shows the
  combined datasets.  These plots illustrate that the predicted \scp
  fall within the expectations of all functional fits at small radii
  and that the probability that the \scp are the result of random
  pairings is extremely low. }
\label{pairhist_50}
\end{center}
\end{figure}

Overlaid on Figures~\ref{pairhist_500} \&~\ref{pairhist_50} are the
pair separation expectations from a linear (solid thin red line) and
power-law (dotted black line) fit to the raw data.  We find that a
power law slope of $\delta=0.53\pm0.01$ provides a best fit for all
three samples.  As illustrated in the two figures, both forms fit well
over all separations, with the linear fit showing a departure from the
data near, and below, $\sim30$ \kpcp.  Continuations of both fits
argue for the existence of LBG pairs with separations smaller than the
imaging resolution.

As mentioned above, interlopers comprise a non-negligible fraction of
colour-selected LBGs.  To estimate the form of the contribution to the
pair separations from random interloper line-of-sight projections and
to estimate the likelihood that the \scp are random projections, we
construct mock catalogues having the physical dimensions of the
images.  We insert galaxies with random locations in the plane of the
sky following the number densities found for each field and the
redshift distributions for each survey.  

Using catalogues of this design is effective at the small separations
studied here because the likelihood of random pairs is low as a result
of the relative low surface density of colour-selected galaxies
($\sim1.7$ arcmin$^{-1}$).  In addition, the enhancement of the number
of random projected close pairs as a result of the spatial clustering
of LBGs is negligible as compared to random line-of-sight projections.
The \zzz colour-selection techniques used here probe $>500$ \mpcp.
The spatial correlation length of \zzz LBGs is $\sim4$ \mpc
\citep{a03,a05,c05} with an average of $\sim1.5$ LBGs over random
integrated over that volume.  Beyond that distance, the clustering
effects of LBGs diminishes and the effects of random projections
become dominant.  Because the projected separations of the \scp are
$\lesssim15$ \kpc, the increase in likelihood that the \scp are random
projections enhanced by the spatial clustering of LBGs is vanishingly
small.

We measure the interloper contribution for each field separately and
combine the results from 1000 runs of the mock catalogues.  The form
of the contamination is shown on Figures~\ref{pairhist_500}
and~\ref{pairhist_50} (dashed thin red line) normalised to the total
number of separations.  From the mock catalogues, we find that the
fraction of random projected pairs with separations $<15$ \kpc
averaged over the 26 fields is $0.0002\pm0.00009$.  As a result, the
\scp are not random projections.

To estimate the interloper contribution to the pair separations, we
minimised a $\chi^2$ fit of a two component distribution that includes
the random catalogue results and the angular clustering of LBGs.  For
LBG angular clustering, we used the conventional form of the angular
correlation function, $\omega (\theta) = A\theta^{-\delta}$.  We first
test the fit by fixing the slope of the angular correlation function
to $\delta=0.57$ as found by \citet{a05} and varying the fraction of
interlopers, $f_I$.  We find the best fit when using an interloper
fraction of $f_I=0.196$ and $f_I=0.195$ for the C05 and S03 samples,
respectively.  The contribution by the LBG angular correlation
function is shown in Figures~\ref{pairhist_500} and~\ref{pairhist_50}
as the thick dashed (light blue) lines with the combined two-component
LBG/random contribution (corrected angular correlation function) shown
by the thick solid (blue) lines.  To help illustrate the effect the
slope of the angular correlation function has on the results, we also
show the LBG contribution when using a slope of $\delta=0.8$ (dashed
green line).  

We re-fit the data letting both parameters vary freely and found best
fits in $\delta$ and $f_I$ within $\sim3$\% of the above values.
Because this type of test has a small dependence on choice of bin
size, we use the fixed $\delta=0.57$ \citep{a05} result in our
assessment of \scp expectations.  The consistency in pair
distributions and parameter fits for both surveys illustrates the
utility of using pair separations to describe the behaviour of LBGs on
small scales.

The number of LBG close pairs predicted by the corrected angular
correlation function provides a good match to the observed pairs on
all scales down to the resolution of the images and becomes
indistinguishable from the uncorrected LBG angular correlation
function expectations at separations smaller than $\sim30$ \kpcp.
This agreement reiterates the negligible contribution of random
projected pairs on small scales.

The widths of our spectroscopic slitlets are $\lesssim8$ \kpcp.
Inspecting Figure~\ref{pairhist_50} at separations of $\lesssim15$
\kpc shows that the expectation from every functional fit to the data
predicts that we should find close pairs serendipitously in our
slitlets.  From the fit of the corrected (and uncorrected) LBG angular
correlation function, we find that both the expected number of \scp
and the expected distribution of SSCpair separations agree well with
that of the observations (shown as the dashed histogram in
Figure~\ref{pairhist_50}).
 
\section{SERENDIPITOUS PAIR EXPECTATIONS FROM SIMULATION}\label{sim}

We test the identification of the \scp against the predictions of LCDM
by analysing a high-resolution hybrid numerical/analytical
cosmological simulation \citep{berrier06}.  The full description of
the simulation and the close pair analysis is presented in a companion
paper, \citet{berrier09}.  Below we describe a few of the relevant
specifics and results.

For large-scale structure, the simulation uses an adaptive refinement
tree N-body code \citep{kravtsov97} that follows the evolution of
$512^3$ particles in a comoving box $120$ \h Mpc on a side.  Numerical
overmerging on small scales is overcome by incorporating the analytic
method of \citet{zentner05} to trace the substructure.  We sample the
simulation at the $z=3$ time-step and use dark matter halos in the
simulation that match the spatial correlation function of the form
$\xi (r) = (r/r_0)^{-\gamma}$ measured for the m$_R\le 25.5$ LBG
samples.  We adopt the spatial correlation length $r_0=3.3\pm0.6$
\citep{c06} using a fixed slope $\gamma=1.6$\footnote{In \citet{c06}
we use several methods to measure and test the correlation function
parameters because of the inherent uncertainties in analysing
$\sim200$ LBGs.  We choose the value we find most reliable for the 140
LBGs used here that is obtained from a maximum likelihood analysis
holding $\gamma$ fixed.} for the C05 survey and $r_0=4.0\pm0.6$ and
$\gamma=1.57\pm0.14$ \citep{a05} for the S03 survey.  We find that
halos with $v_{circ}\ge140$ \kms in the simulation yield a correlation
length of $r_0=3.9\pm1.5$ and slope of $\gamma=1.6\pm0.3$.  These
halos correspond to a number density of $n_{LBG}=0.016~h^{3}$
Mpc$^{-3}$, comoving, and halo mass $M_{Halo}\gtrsim10^{11.3} M_\odot$
($\langle M_{Halo}\rangle\sim10^{11.6} M_\odot$) which are in very
good agreement with the corrected volumetric density from the
photometric sample and the mass inferred from clustering statistics
\citep[$\sim10^{11.5} M_\odot$,][]{a03,a05,c06}.

The simulation is then analysed in a manner identical to the
observations using (1) the constraints from the seeing FWHM and
centroid accuracy from the images, (2) the MOS spectroscopic slitlet
dimensions of our survey, (3) the spectral resolution, and (4) the
uncertainties in determining the systemic LBG spectroscopic redshifts.
We compute the number of pairs that fall in mock slitlets with random
orientations that are resolvable in the images and that have velocity
offsets that would be detectable in LRIS spectroscopy.  We find an
expected fraction of $0.052\pm0.002$, or an average of $7.1$ LBGs
($3-4$ pairs) that is to be compared to the $0.069\pm0.023$ fraction
found from the analysis of this survey.  This close agreement helps
reinforce the interpretation that the \scp are indeed close \zzz LBG
pairs as a similar number are expected from LCDM to be found
serendipitously in our survey.  We find that the expected fraction of
\scp from the simulation is rather insensitive (changes by a few
percent) to a variation of the LBG correlation parameters within the
range of their uncertainties.


\section{SERENDIPITOUS LY-$\alpha$ EMITTERS}\label{lae}

\begin{figure}
\begin{center}
\scalebox{0.39}[0.37]{\rotatebox{90}{\includegraphics{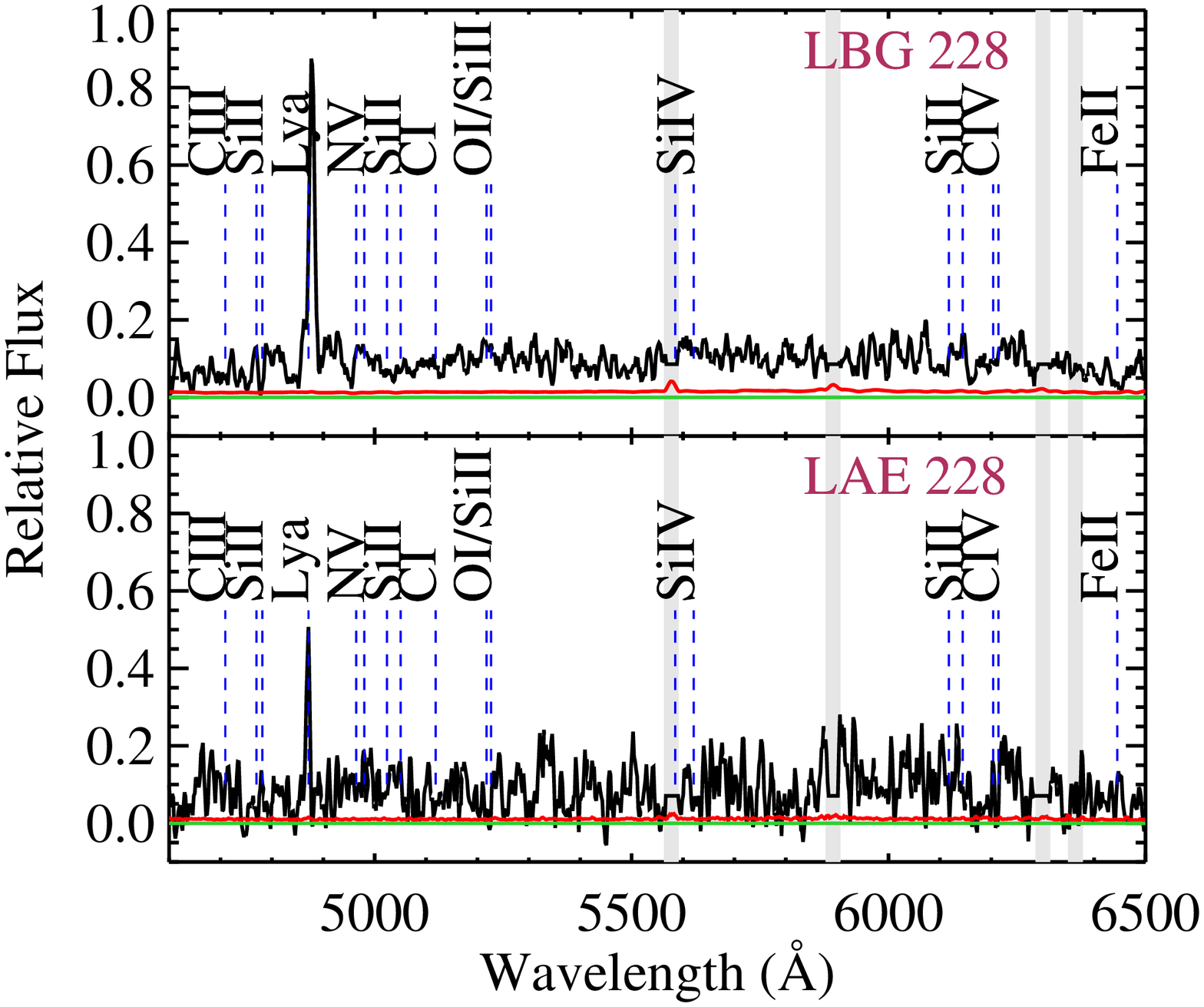}}}
\scalebox{0.39}[0.37]{\rotatebox{90}{\includegraphics{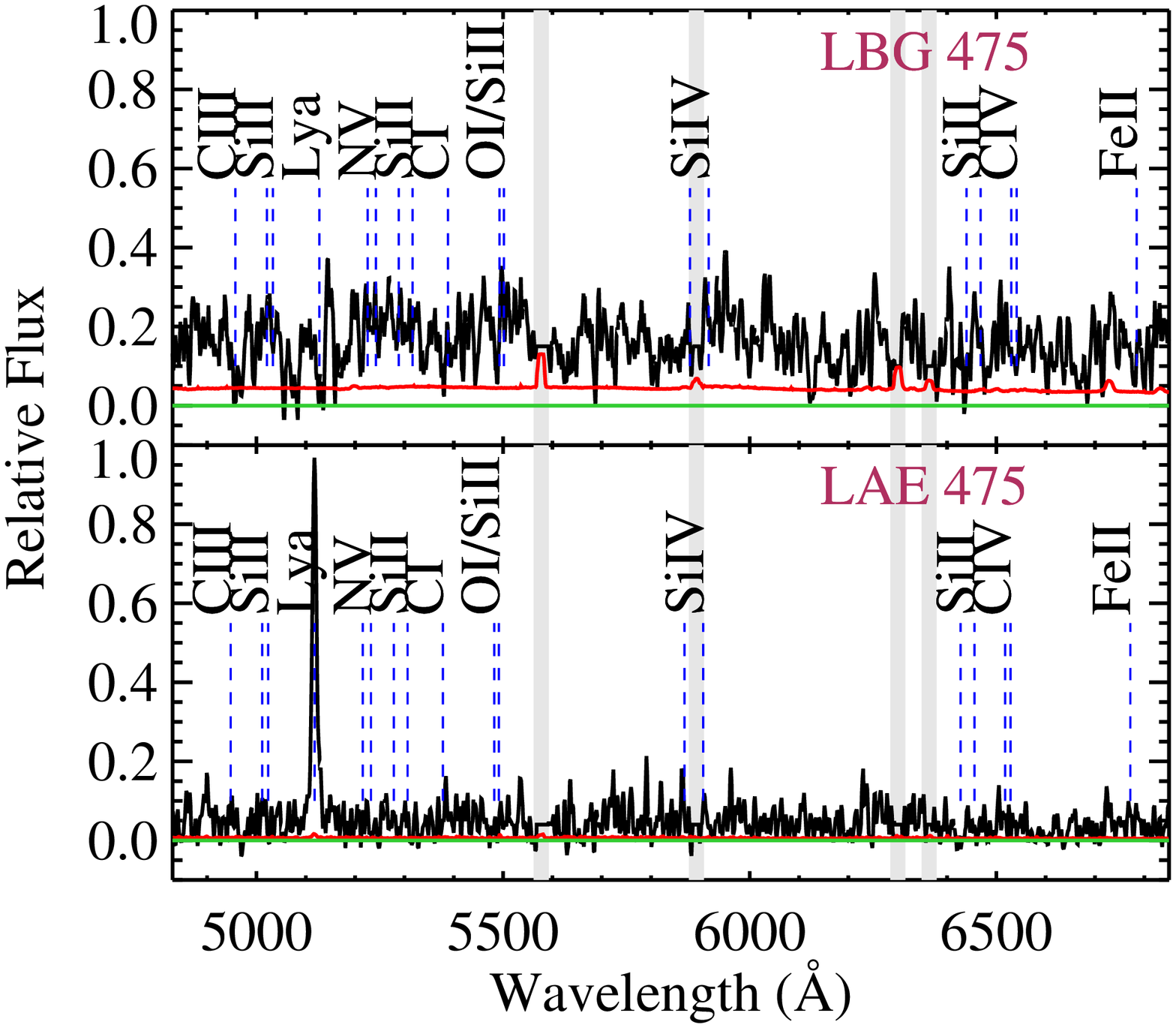}}}
\scalebox{0.72}[0.64]{\includegraphics{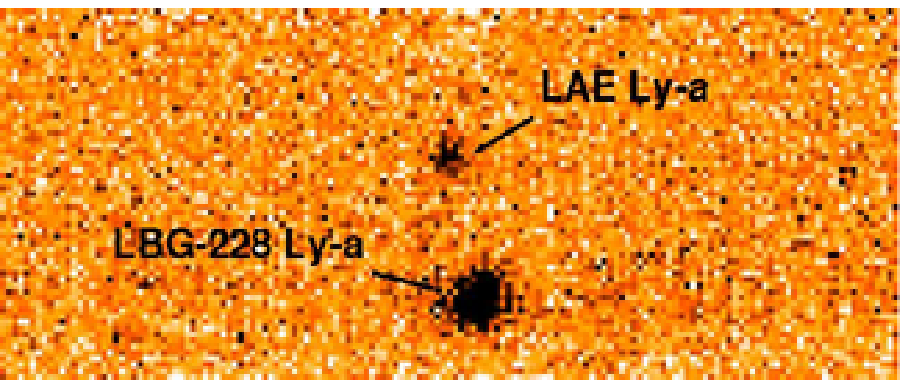}}
\scalebox{0.72}[0.68]{\includegraphics{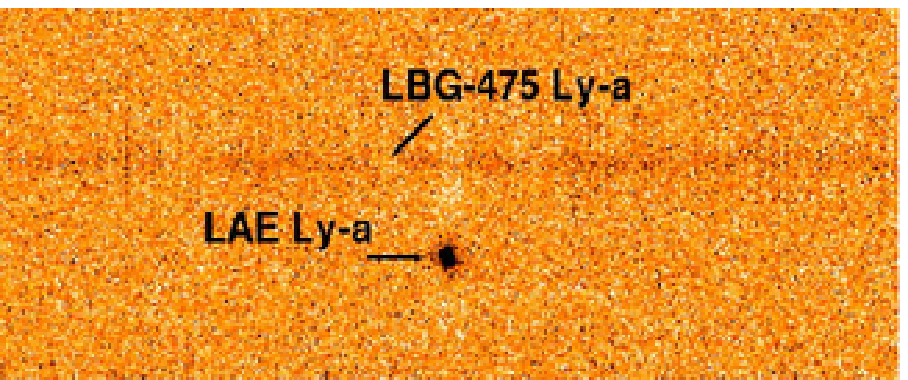}}
\caption
{\small Spectra of the LAEs and close LBGs detected in the survey.
  {\bf Top split panel:} 1-D spectra of LBG-228 and LAE-228 showing
  the \lya emission and continua of both objects. {\bf Centre split
  panel:} 1-D spectra of LBG-475 and LAE-475.  LAE-475 show strong
  \lya emission, however, LBG-475 is considered a probable LBG because
  of the low S/N and inconclusive nature of the ISM line fit.  {\bf
  Bottom two panels:} The 2-D spectra for pairs 228 and 475.  The
  position of the \lya features and continua for pair 228 and 475 are
  indicated and have projected separations of $22.7$ and $40.1$ \kpcp,
  respectively. Although they are within the close pair separation
  criteria, they are not included in the analysis because the LAEs do
  not meet the m$_R<25.5$ criterion. }
\label{lae1D}
\end{center}
\end{figure}

\lya emitters (LAEs) are defined here as objects displaying detectable
\lya emission but have continua fainter than $R=25.5$ and therefore
are not candidates for inclusion in the LBG spectroscopic sample.  We
discovered two serendipitous $z\sim3$ LAEs in the MOS slitlets of the
full C05 LBG sample.  This is approximately the number we would expect
to fall serendipitously into our slitlets, when we consider (1) the
depth of our spectroscopy, (2) the $\sim1$ arcmin$^2$ solid angle
subtended by our slitlets, and (3) the \zzz LAE densities reported by
\citet{gaw07} and \citet{grove09} when assumed to be constant over the
redshift path probed.  The detected objects are LAE-228 at $z=3.007$
in field PC1643+4631A and LAE-475 at $z=3.209$ in the field
BRI1013+0035.  Figure~\ref{lae1D} presents their 1-D and 2-D spectra.
Both redshifts are measured from their \lya emission peaks and exhibit
\lya EW of 48\AA~and 140\AA, respectively.

Interestingly, LAE-228 was found as a spectroscopic close pair to the
spectroscopically confirmed LBG-228 with $z=3.008$.  The projected
separation between the two systems is $22.7$ kpc and they exhibit a
$\Delta v_{Ly\alpha}= 450$ \kmsp.  In addition, LAE-475 was found near
LBG-475 with $z=3.218$ having a projected separation of $40.1$ kpc and
a \lya velocity offset of $650$ \kmsp.  Although both systems could be
considered close pairs, neither is included in the analysis in this
paper because the LAEs do not meet the m$_R\ge25.5$ criterion and
because LBG-475 is considered a ``probable'' LBG \citep[category 4
in][]{c06} from the lower S/N and quality of the spectral-line fit.

LAEs are believed to be systems with lower mass and lower star
formation rates than that of the LBG population
\citep[$\sim10^{10.5-11} M_\odot$ and $\le 10 M_\odot$ yr$^{-1}$;
e.g.,][]{gaw07} that are undergoing a recent star forming episode.
The fact that the two LAEs detected are close pairs to one, and likely
both, LBGs suggests that a fraction of the \lya emission, and
therefore the detectability of LAEs, may result from interactions.


\section{\lya emission and close pairs}\label{ecp}

In our analysis, we see dominant \lya in emission from all galaxies of
the \scp and a prevalence for dominant \lya emission for LBGs having
close photometric companions (SSCpair candidates; see
Figures~\ref{phot_pairs} \&~\ref{Rband_2Dnon}).  We consider the
frequency of the appearance of \lya in emission for the \scp and the
SSCpair candidates.  The fraction of eLBGs in our spectroscopic sample
of 140 LBGs is $0.55$.  This fraction includes LBGs exhibiting purely
emission with undetectable absorption, to those with a complex profile
that show \lya absorption but have strong enough emission to result in
net \lya emission EW.  By our definition, the larger spectroscopic
sample of S03 finds a comparable eLBG fraction of $0.51$.  We remark
that the low S/N, low-resolution LBG continua and complex \lya
features near zero net EW, where the peak of the EW distribution is
located, may produce minor differences between the \citet{aes03}
analysis of the S03 data and ours.  These differences may contribute
toward the small discrepancy in the two eLBG/aLBG ratios, however, do
not affect the results below for LBGs with $<15$ \kpc separations as
none of these exhibit such \lya profiles.  In the probability
calculations below, we use the eLBG fraction from our survey
exclusively.  The eLBG fraction found by \citet{aes03} analysis of the
S03 data would result in lower probabilities.

The binomial probability of observing all 10 SSCpair galaxies with
dominant \lya emission and zero galaxies with dominant \lya in
absorption by chance is $p=0.0025$.  The six spectroscopic LBG
components of the SSCpair candidates all show dominant \lya in
emission but were not included in the \scp analysis because the bulk
of the secondary component did not enter the slitlet
(Figure~\ref{phot_pairs} and Figure~\ref{Rband_2Dnon}).  Combined, the
binomial probability that all 16 galaxies are eLBGs and zero are aLBGs
is $p=7.0\times10^{-5}$.  If the emission of these systems was a
result of star forming clumps or single LBGs, we would expect to
observe an approximately equal fraction (or $\sim45$\% for this
survey) of regions that are dominated by \lya absorption and we
observe none.  As a reminder, systems exhibiting \lya absorption
(aLBGs) were searched for thoroughly in the data (\S\ref{spectro}).

\begin{figure}
\begin{center}
\scalebox{0.45}[0.40]{\includegraphics{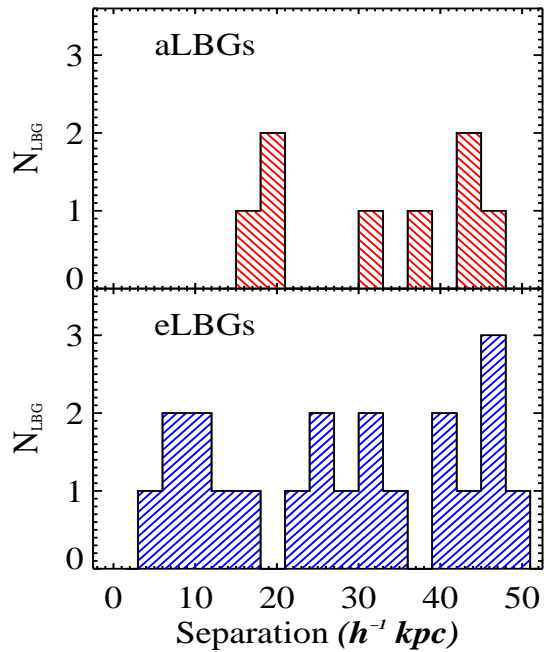}}
\caption
{\small Histograms of the \lya behaviour of spectroscopic LBGs with
  close photometric companions.  Pair separations where the
  spectroscopic component exhibits dominant \lya absorption indicated
  by the red (grey) back-hatched histogram (upper panel) and dominant
  \lya emission are indicated by the blue (black) forward-hatched
  histogram (lower panel).  The \scp are not included.  All SSCpair
  components have separations of $<15$ \kpc and show \lya emission.
  Pairs shown with $<15$ \kpc separations include the SSCpair
  candidates. }
\label{bluered}
\end{center}
\end{figure}

When examining the \lya features of spectroscopic LBGs having
photometric companions that meet our colour criteria and have
projected separations between $15-50$ \kpcp, with the caveat that some
of these may not be physical pairs, we find $11$ eLBGs and $7$ aLBGs
($p=0.17$).  In addition, if we include all spectroscopically
confirmed LBGs with photometric close pairs that have relaxed colour
criteria (see C05) and that have projected separations between $15-50$
\kpc (those with $<15$ \kpc are included in the SSCpair candidates),
we find a total of $15$ eLBGs and $8$ aLBGs ($p=0.11$).  These two
measurements suggest an overabundance of eLBGs and, combined with the
abundance of \lya emission in the \scp and SSCpair candidates,
indicate an overabundance that diminishes with increasing radius.
Figure~\ref{bluered} plots the number of eLBGs and aLBGs with
photometric companions with separation.  Finally, LBG-228 with close
pair LAE-228 ($22.7$ \kpcp) exhibits \lya in emission.  The tentative
LBG-475 with close pair LAE-475 ($40.1$ \kpcp) shows weak evidence of
complex \lya absorption and emission in the low S/N spectrum, but a
proper EW analysis is not possible at this time.

We posit that the overabundance of eLBGs at small separations
($\lesssim15$ \kpcp) is caused, in part, by induced star formation and
the dispersal of gas and dust related interactions.  Object D3 in the
sample of \citet{aes03} provides further support in that both galaxies
in the identified close pair ($\sim15$ \kpc separation) exhibit
dominant \lya emission.  The extended \lya emission of D3, \lya blob
nature, and potential fraction of escaping Lyman continuum photons
reinforces this scenario.  Finally, the trends of \lya emission with
morphological parameters are consistent with a picture where a
non-negligible fraction of LBGs produce double-peak \lya emission when
merging their ground-based 1-D spectra as a result of triggered or
more revealed \lya emission.

The statistics presented here can be quickly improved from the
analysis of other similar existing and future datasets.  Nevertheless,
the \lya behaviour of the \scp appears real.  Such behaviour would
suggest that although star formation is ubiquitous at high redshift,
caused in part by the high gas fraction in most systems, and generates
\lya emission, close pairs have additional \lya emission and UV flux
as a result of an interaction.  The enhanced \lya emission from
interactions modifies the ratio of eLBGs and increases the fraction of
detectable eLBGs for a given criteria whereas without an interaction
they would remain either too faint or too red.  Such behaviour would
bias any colour-selection technique.  Similarly, interactions may make
LAEs detectable whereas without an interaction they would remain too
faint to be detected from their continuum alone.

We find $\sim7$\% of the galaxies in our spectroscopic sample
identified as interacting pairs.  If the higher fraction of \lya
double-peak systems in the higher S/N, higher resolution sample of
\citet{aes06} prove to be interacting pairs, because both samples are
serendipitous detections, the true fraction of interacting LBGs at
\zzz may be $\gtrsim30$\%.  If borne out, the double \lya
emission/double morphology observable in ground-based data may provide
a simple means to identify a significant fraction of interacting
galaxies and help to constrain the merger rate at high redshift.  The
relationship between \lya emission and pair separation will be fully
explored in a future paper.  The evidence is presented here to
establish a link between the presence of \lya in emission and the
identification of interacting pairs.

In conclusion, the above picture and the observations of merged 1-D
double \lya emission profiles in ground-based data as markers of this
process are supported by: (1) the detection of \lya in emission from
every component of the \scp and the spectroscopic components of all
six SSCpair candidates, (2) an indication of an overabundance of eLBGs
in spectro-photometric $\le50$ \kpc close pairs, (3) the trends of
\lya emission EW with nucleation, colour, and separation
(size/multiplicity) of \citet{law07}, (4) the plausibility from \lya
and ISM feature analysis that the double peak systems in the work of
\citet{aes06} are close and/or interacting pairs, (5) the fact that
both galaxies in the identified close pair D3 exhibit \lya in emission
and the identification of this system as an extended \lya blob, (6)
the potential Lyman continuum escaping photons from D3 and double-peak
\lya emission system C49, and (7) the detection of one, and
potentially two, LAEs within $\le50$ \kpc of spectroscopic LBGs in
this survey.


\section{SUMMARY}\label{sum}

A systematic search of deep Keck imaging and spectroscopy of 140 \zzz
LBGs in nine separate fields has uncovered five serendipitous
spectroscopic close pairs (\scpp) with separations of $<15$
\kpcp. Below, we summarise our findings and the implications of the
imaging and spectroscopic data from this survey, the photometric data
from the survey of S03, and the results from an analysis of a
carefully matched high-resolution cosmological hybrid simulation.\\

\noindent (1) We find evidence in the imaging and spectroscopic
    observations that the \scp are two distinct and likely interacting
    systems.

\begin{enumerate}

\item The \scp exhibit two flux peaks with $\sim3-13$ \kpc projected
  separations in the broadband images.  The VRI filters probe the
  continua of O and B stars at \zzz and the SSCpair components exhibit
  luminosities that are typical of \zzz LBGs.  In addition, morphology
  of the \scp is consistent with the expectations of low to
  intermediate redshift UVLGs projected to \zzzp.  UVLGs are in every
  way analogues to LBGs having similar mass, gas content, UV colours,
  metallicity, half-light radii, and star formation rates.
  Mergers/interactions are determined to be the main star-forming
  mechanism for UVLGs in the sample of O08.

\item Each component of the \scp exhibits detectable \lya emission and
  a corresponding 2-D continuum.  The continua of the \scp are closely
  spaced or overlapping.  When merging the 1D spectra, each SSCpair
  shows a double-peak \lya emission profile which is therefore
  different in nature than those modeled for a static or expanding
  shell of gas driven by galactic-scale outflows.  The \lya emission
  peaks and continua in the 2D spectra are spatially offset and
  exhibit a one-to-one spatial correspondence with continuum flux
  peaks in the images.

\item The velocity offsets between the two \lya emission peaks of the
  \scp (before and after angular separation corrections) are not
  representative of a random sampling of \lya offsets pulled from the
  LBG population if the two features are generated by two sources
  within single systems.  Instead, the \lya offsets are in agreement
  with single \lya emission from two separate systems.

\item We see evidence for two sets of ISM absorption lines (at a lower
  confidence level) in each merged \scp spectrum having relative
  velocities inconsistent with the expectations of the general LBG
  population but consistent with two separate systems.  In addition,
  the velocity differences between the \lya and ISM features assigned
  to each component of the \scpp, under the assumption that they are
  distinct systems, is consistent with typical LBG values.

\item The high-S/N, higher-resolution \zzz LBG spectroscopic sample of
  \citet{aes06} report that four of the nine eLBGs in their sample of
  14 systems display double-peak \lya emission.  Analysis of the
  reported values finds the data for the three double-peak systems and
  one identified double system to be consistent with the \scp
  observations.  Both samples show \lya and potential ISM offsets and
  \lya-ISM velocity differences that are less consistent with a single
  system and more consistent with typical values for two interacting
  \zzz galaxies.

\item The morphology and \lya features of the \scp are consistent with
  the reported relationships of \citet{law07} between space-based
  morphological parameters (Gini, multiplicity, size, and colour) and
  \lya emission EW of \zzz LBGs.  An interpretation of the these
  relationships points to a picture where single and double-nuclei
  LBGs are \lya emitting LBGs, with a significant fraction being
  interacting systems.

\end{enumerate}

\noindent (2) We examine the distribution of all close pairs in the
data to determine the expectation of serendipitous pairs in our
spectroscopic slitlets.  We find that an angular correlation function
of the form $\omega(\theta) = A\theta^{-\delta}$ with slope
$\delta=0.57$ provides an excellent fit to close pair counts for both
surveys tested.  The fit incorporates contamination by random
interlopers determined from mock catalogues matched to the volume and
behaviour of the data.  The slope of the fit to the close pair counts
is identical to, or in very close agreement with, the \zzz LBG values
in the literature derived from angular and spatial clustering
statistics.  Extrapolation of the fit to below the resolution limit of
the photometric data ($\lesssim10$ \kpcp) predicts the existence of
\scp with a similar number and separation distribution to that
observed.

\noindent (3) We analyse a high-resolution cosmological hybrid
numerical and analytical simulation to assess the expectation of the
\scp from the predictions of LCDM.  We carefully matched the
simulation to the number density, spatial correlation function,
spectral and imaging resolutions of the data, and the dimensions of
the slitlets.  The analysis finds a prediction of $\sim4$ pairs to
fall serendipitously in the multi-object slitlets of our survey, in
good agreement with the observations.

\noindent (4) We present two LAEs discovered serendipitously in the
MOS spectroscopic slitlets which are expected from LAE densities at
\zzz and the solid angle subtended by our slitlets.  However, both
LAEs are $\le50$ \kpc close projected pairs to LBGs.  We confirm one
of the LAEs to lay at the redshift of a nearby LBG and is a bona-fide
close pair ($22.7$ \kpcp), whereas the second is only a probable close
pair because of the low S/N of the LBG spectrum.  In addition, the LBG
in the bona-fide close pair exhibits \lya in emission.

\noindent (5) We uncover a relationship between \lya emission and pair
separation.  We find an overabundance of LBGs with separations of
$<50$ \kpc exhibiting \lya in emission as compared to the full sample.
Moreover, without exception, {\it all galaxies in the \scp and all
SSCpair candidates ($\lesssim15$ \kpc) have dominant \lya emission}.

The presence of \lya in all \scp and \scp candidates supports the
interpretation that the \scp are interacting LBGs under the assumption
that interaction may disperse gas and dust and trigger observable star
formation.  Because of the close proximity of the galaxies in the data
to each other, the double \lya emission/double morphology of the \scp
provides a useful and consistent diagnostic to help detect interacting
close pairs from $2\lesssim z\lesssim 7$ in similarly obtained
existing and future optical ground-based data.


\section*{Acknowledgments}

The authors are supported by the Center for Cosmology at the
University of California, Irvine.  J. C. gratefully acknowledges
generous support by Gary McCue.  The authors wish to recognise and
acknowledge the very significant cultural role and reverence that the
summit of Mauna Kea has always had within the indigenous Hawaiian
community.  We are most fortunate to have the opportunity to conduct
observations from this mountain.

\bsp

\label{lastpage}


\begin{thebibliography}{99}
\bibitem[\protect\citeauthoryear{Adelberger et al.}{2003}]{a03}
  Adelberger, K. L., Steidel, C. C., Shapley, A. E., \& Pettini,
  M. 2003, ApJ, 584, 45
\bibitem[\protect\citeauthoryear{Adelberger et al.}{2005}]{a05}
  Adelberger, K. L., Steidel, C. C., Pettini, M., Shapley, A. E.,
  Reddy, N. A., \& Erb, D. K. 2005, ApJ, 619, 697
\bibitem[\protect\citeauthoryear{Barton et al.}{2000}]{barton00}
  Barton, E. J., Geller, M. J., \& Kenyon, S. J. 2000, ApJ, 530, 660
\bibitem[\protect\citeauthoryear{Barton et al.}{2003}]{barton03}
  Barton, E. J., Geller, M. J., \& Kenyon, S. J. 2003, ApJ, 582, 668
\bibitem[\protect\citeauthoryear{Berrier et al.}{2006}]{berrier06}
  Berrier, J. C., Bullock, J. S., Zentner, A. R., Guenther, H. D.,
  Barton, E. J., Kravtsov, A. V., \& Wechsler, R. H. 2006, ApJ, 652,
  56
\bibitem[\protect\citeauthoryear{Berrier et al.}{2009}]{berrier09}
  Berrier, J. C., Cooke, J., Bullock, J. S., Zentner, A. R., Guenther,
  H., Barton, E. J., 2009 ApJ, submitted
\bibitem[\protect\citeauthoryear{Bertin \& Arnouts}{1996}]{ba96}
  Bertin, E. \& Arnouts, S. 1996, A\&AS, 117, 393
\bibitem[\protect\citeauthoryear{Cooke et al.}{2005}]{c05} Cooke, J.,
  Wolfe, A. M., Gawiser, E., \& Prochaska, J. X. 2005, ApJ, 621, 596
\bibitem[\protect\citeauthoryear{Cole et al.}{2000}]{cole00} Cole, S.,
  Lacey, C.~G., Baugh, C.~M., \& Frenk, C.~S.\ 2000, MNRAS, 319, 168
\bibitem[\protect\citeauthoryear{Conselice et al.}{2003}]{conselice03}
  Conselice, C.~J., Bershady, M.~A., Dickinson, M., \& Papovich, C.\
  2003, AJ, 126, 1183
\bibitem[\protect\citeauthoryear{Cooke et al.}{2006}]{c06} Cooke, J.,
  Wolfe, A. M., Prochaska, J. X., \& Gawiser, E. 2006 ApJ, 652, 994
\bibitem[\protect\citeauthoryear{Cooke et al.}{2008}]{c08} Cooke, J.,
  Barton, E.~J., Bullock, J.~S., Stewart, K.~R., \& Wolfe, A.~M.\
  2008, ApJL, 681, L57
\bibitem[\protect\citeauthoryear{Erb et al.}{2003}]{erb03} Erb, D.~K.,
  Shapley, A.~E., Steidel, C.~C., Pettini, M., Adelberger, K.~L.,
  Hunt, M.~P., Moorwood, A.~F.~M., \& Cuby, J.-G.\ 2003, ApJ, 591, 101
\bibitem[\protect\citeauthoryear{Erb et al.}{2006}]{erb06} Erb, D.~K.,
  Steidel, C.~C., Shapley, A.~E., Pettini, M., Reddy, N.~A., \&
  Adelberger, K.~L.\ 2006, ApJ, 646, 107
\bibitem[\protect\citeauthoryear{Ferguson et al.}{2004}]{ferg04}
  Ferguson, H.~C., et al.\ 2004, ApJL, 600, L107
\bibitem[\protect\citeauthoryear{F{\"o}rster Schreiber et
  al.}{2006}]{fs06} F{\"o}rster Schreiber, N.~M., et al.\ 2006, ApJ,
  645, 1062
\bibitem[\protect\citeauthoryear{Fukugita et al.}{1996}]{f96}
  Fukugita, M., Ichikawa, T., Gunn, J. E., Doi, M., Shimasaku, K., \&
  Schneider, D. P. 1996, AJ, 111, 1748
\bibitem[\protect\citeauthoryear{Gardner et al.}{2000}]{g00} Gardner,
  J. P., Brown, T. M., \& Ferguson, H. C. 2000, ApJ, 542L, 79
\bibitem[\protect\citeauthoryear{Gawiser et al.}{2007}]{gaw07}
  Gawiser, E., et al.\ 2007, ApJ, 671, 278
\bibitem[\protect\citeauthoryear{Genzel et al.}{2008}]{genzel08}
  Genzel, R., et al.\ 2008, ApJ, 687, 59
\bibitem[\protect\citeauthoryear{Grazian et al.}{2007}]{grazian07}
  Grazian, A., et al.\ 2007, A\&A, 465, 393 
\bibitem[\protect\citeauthoryear{Grimes et al.}{2007}]{grimes07}
  Grimes, J.~P., et al.\ 2007, ApJ, 668, 891
\bibitem[\protect\citeauthoryear{Grove et al.}{2009}]{grove09} Grove,
  L.~F., Fynbo, J.~P.~U., Ledoux, C., Limousin, M., M{\o}ller, P.,
  Nilsson, K.~K., \& Thomsen, B.\ 2009, A\&A, 497, 689
\bibitem[\protect\citeauthoryear{Heckman}{2002}]{heckman02} Heckman,
  T.~M.\ 2002, Extragalactic Gas at Low Redshift, 254, 292
\bibitem[\protect\citeauthoryear{Heckman et al.}{2005}]{heckman05}
  Heckman, T.~M., et al.\ 2005, ApJL, 619, L35
\bibitem[\protect\citeauthoryear{Kartaltepe et al.}{2007}]{kart07}
  Kartaltepe, J.~S., et al.\ 2007, ApJS, 172, 320
\bibitem[Keel et al.(1985)]{keel85} Keel, W.~C., Kennicutt, R.~C.,
  Jr., Hummel, E., \& van der Hulst, J.~M.\ 1985, AJ, 90, 708
\bibitem[\protect\citeauthoryear{Kells et al.}{1998}]{k98} Kells, W.,
  Dressler, A., Sivaramakrishnan, A., Carr, D., Koch, E., Epps, H.,
  Hilyard, D., \& Pardeilhan, G. 1998, PASP, 110, 1487
\bibitem[\protect\citeauthoryear{Kravtsov et al.}{1997}]{kravtsov97}
  Kravtsov, A. V., Klypin, A. A., \& Khokhlov, A. M. ~1997, ApJS, 111,
  73
\bibitem[\protect\citeauthoryear{Lambdas et
  al.}{2003}]{lambdas03}Lambas, D. G., Tissera, P. B., Sol Alonso, M.,
  \& Coldwell, G. 2003, MNRAS, 346, 1189
\bibitem[Larson \& Tinsley(1978)]{larson78} Larson, R.~B., \& Tinsley,
  B.~M.\ 1978, ApJ, 219, 46
\bibitem[\protect\citeauthoryear{Law et al.}{2007}]{law07} Law, D. R.,
  Steidel, C. C., Erb, D. K., Pettini, M., Reddy, N. A., Shapley,
  A. E., Adelberger, K. L., \& Simenc, D. J.  2007, ApJ, 656, 1
\bibitem[\protect\citeauthoryear{Law et al.}{2009}]{law09} Law, D.~R.,
  Steidel, C.~C., Erb, D.~K., Larkin, J.~E., Pettini, M., Shapley,
  A.~E., \& Wright, S.~A.\ 2009, ApJ, 697, 2057
\bibitem[\protect\citeauthoryear{Lin et al.}{2004}]{lin04} Lin, L., et
  al.\ 2004, ApJL, 617, L9
\bibitem[\protect\citeauthoryear{Lin et al.}{2007}]{lin07} Lin, L., et
  al.\ 2007, ApJL, 660, L51
\bibitem[\protect\citeauthoryear{Lin et al.}{2008}]{lin08} Lin, L., et
  al. 2008, ApJ, 681, 232
\bibitem[\protect\citeauthoryear{Lotz et al.}{2006}]{lotz06} Lotz, J.,
  Madau, P., Giavalisco, M., Primack, J., \& Ferguson, H. 2006, ApJ,
  636, 592
\bibitem[\protect\citeauthoryear{Marchesini et al.}{2007}]{marchesini07}
  Marchesini, D., et al.\ 2007, ApJ, 656, 42 
\bibitem[\protect\citeauthoryear{Mas-Hesse et al.}{2003}]{m03}
  Mas-Hesse, J.~M., Kunth, D., Tenorio-Tagle, G., Leitherer, C.,
  Terlevich, R.~J., \& Terlevich, E.\ 2003, ApJ, 598, 858
\bibitem[\protect\citeauthoryear{Matsuda et al.}{2004}]{matsuda04}
  Matsuda, Y., et al.\ 2004, AJ, 128, 569
\bibitem[\protect\citeauthoryear{Mihos \& Hernquist}{1996}]{mihos96}
  Mihos, J.~C., \& Hernquist, L.\ 1996, ApJ, 464, 641
\bibitem[\protect\citeauthoryear{McCarthy et al.}{1998}]{mccarthy98}
  McCarthy, J.~K., et al.\ 1998, SPIE, 3355, 81
\bibitem[\protect\citeauthoryear{Nikolic et
  al.}{2004}]{nikolic04}Nikolic, B., Cullen, H., \& Alexander,
  P. 2004, MNRAS, 355, 874
\bibitem[\protect\citeauthoryear{Oke et al.}{1995}]{o95} Oke, J. B.,
  Cohen, J. G., Carr, M., Cromer, J., Dingizian, A., Harris, F. H.,
  Labrecque, S., Lucinio, R., Schaal, W., Epps, H., \& Miller, J.
  1995, PASP, 107, 375
\bibitem[\protect\citeauthoryear{Overzier et al.}{2008}]{overzier08}
  Overzier, R.~A., et al.\ 2008, ApJ, 677, 37
\bibitem[\protect\citeauthoryear{Papovich et al.}{2001}]{pap01}
  Papovich, C., Dickinson, M., \& Ferguson, H.~C.\ 2001, ApJ, 559, 620
\bibitem[\protect\citeauthoryear{Patton et al.}{2000}]{patton00}
  Patton, D.~R., Carlberg, R.~G., Marzke, R.~O., Pritchet, C.~J., da
  Costa, L.~N., \& Pellegrini, P.~S.\ 2000, ApJ, 536, 153
\bibitem[\protect\citeauthoryear{Patton et al.}{2002}]{patton02}
  Patton, D.~R., et al.\ 2002, ApJ, 565, 208
\bibitem[\protect\citeauthoryear{Rafelski et al.}{2009}]{rafelski09}
  Rafelski, M., Wolfe, A. M., Cooke, J., Chen, H.-W., Armandroff,
  T. E., Wirth, G. D.\ 2009, in preparation
\bibitem[\protect\citeauthoryear{Reddy et al.}{2005}]{reddy05} Reddy,
  N.~A., Erb, D.~K., Steidel, C.~C., Shapley, A.~E., Adelberger,
  K.~L., \& Pettini, M.\ 2005, ApJ, 633, 748
\bibitem[\protect\citeauthoryear{Scannapieco \&
  Tissera}{2003}]{scan03} Scannapieco, C., \& Tissera, P.~B.\ 2003,
  MNRAS, 338, 880
\bibitem[\protect\citeauthoryear{Shapiro et al.}{2008}]{shapiro08}
  Shapiro, K.~L., et al.\ 2008, ApJ, 682, 231
\bibitem[\protect\citeauthoryear{Shapley et al.}{2001}]{aes01}
  Shapley, A. E., Steidel, C. C., Adelberger, K. L., Dickinson, M.,
  Giavalisco, M., \& Pettini, M. 2001, ApJ, 562, 95
\bibitem[\protect\citeauthoryear{Shapley et al.}{2003}]{aes03}
  Shapley, A. E., Steidel, C. C., Adelberger, K. L., \& Pettini,
  M. 2003, ApJ, 588, 65
\bibitem[\protect\citeauthoryear{Shapley et al.}{2006}]{aes06}
  Shapley, A. E., Steidel, C. C., Pettini, M., Adelberger, K. L., \&
  Erb, D. K. 2006, ApJ, 651, 688
\bibitem[\protect\citeauthoryear{Steidel et al.}{1996}]{s96} Steidel,
  C.~C., Giavalisco, M., Pettini, M., Dickinson, M., \& Adelberger,
  K.~L.\ 1996, ApJL, 462, L17
\bibitem[\protect\citeauthoryear{Steidel et al.}{2003}]{s03} Steidel,
  C. C., Adelberger, K. L., Shapley, A. E., Pettini, M., Dickinson,
  M., \& Giavalisco, M. 2003, ApJ, 592, 728
\bibitem[\protect\citeauthoryear{Steinmetz \&
  Navarro}{2002}]{steinmetz02} Steinmetz, M., \& Navarro, J.~F.\ 2002,
  New Astronomy, 7, 155
\bibitem[\protect\citeauthoryear{Tenorio-Tagle et al.}{1999}]{tt99},
  Tenorio-Tagle, G., Silich, S. A., Kunth, D., et al. 1999, MNRAS,
  309, 332
\bibitem[\protect\citeauthoryear{Vanzella et al.}{2008}]{vanzella08}
  Vanzella, E., et al.\ 2008, A\&A, 487, 83
\bibitem[\protect\citeauthoryear{Verhamme et al.}{2006}]{verhamme06}
  Verhamme, A., Schaerer, D., \& Maselli, A.\ 2006, A\&A, 460, 397
\bibitem[\protect\citeauthoryear{Verhamme et al.}{2008}]{verhamme08}
  Verhamme, A., Schaerer, D., Atek, H., \& Tapken, C.\ 2008, A\&A,
  491, 89
\bibitem[\protect\citeauthoryear{Wright et al.}{2009}]{wright09}
  Wright, S.~A., Larkin, J.~E., Law, D.~R., Steidel, C.~C., Shapley,
  A.~E., \& Erb, D.~K.\ 2009, ApJ, 699, 421
\bibitem[\protect\citeauthoryear{Zentner et al.}{2005}]{zentner05}
  Zentner, A. R., Berlind, A. A., Bullock, J. S., Kravtsov, A. V., \&
  Weschler, R. H. ~2005, ApJ, 624, 505
\end{thebibliography}
\end{document}